\documentclass{aa}  

\usepackage{graphicx}
\usepackage{txfonts}

\newcommand{\be}{\begin{equation}}
\newcommand{\ee}{\end{equation}}
\def\ltsima{$\; \buildrel < \over \sim \;$}
\def\lsim{\lower.5ex\hbox{\ltsima}}
\def\gtsima{$\; \buildrel > \over \sim \;$}\def\gsim{\lower.5ex\hbox{\gtsima}}

\begin{document}

\title{When GRB afterglows get softer, hard components come into play  }
\author{A. Moretti\inst{1}, R. Margutti\inst{1,2}, F. Pasotti\inst{1,2}, A.P. Beardmore\inst{3}, S. Campana\inst{1}, G. Chincarini\inst{2,1},
S. Covino\inst{1}, O. Godet\inst{3}, C. Guidorzi\inst{2,1},  J.P. Osborne\inst{3}, P. Romano\inst{2,1}, G. Tagliaferri\inst{1}} 
\offprints{alberto.moretti@brera.inaf.it}

\institute
{
INAF, Osservatorio Astronomico di Brera, Via E. Bianchi 46, I-23807, Merate (LC), Italy 
\and Universit\`a degli Studi di Milano-Bicocca, Dipartimento di Fisica, Piazza delle Scienze 3, I-20126 Milano, Italy 
\and  University of Leicester, LE1 7RH, UK
} 
\date{Received ; accepted }
\date{Received ; accepted }
\titlerunning{When GRB get softer ... }
\authorrunning{Moretti et al.}  
\abstract{} 
{We aim to investigate the ability of simple spectral
models to describe the GRB early afterglow emission.}
{We performed a time resolved spectral analysis of a bright GRB sample
detected by the Swift Burst Alert Telescope and promptly observed by
the Swift X--ray Telescope,with spectroscopically measured redshift in
the period April 2005 -- January 2007.  The sample consists of 22 GRBs
and a total of 214 spectra. We restricted our analysis to the softest
spectra sub--sample which consists of 13 spectra with photon index $>$
3.}
{In this sample we found that four spectra, belonging to GRB060502A,
GRB060729, GRB060904B, GRB061110A prompt--afterglow transition phase,
cannot be modeled neither by a single power law nor by the Band model.
Instead we find that the data present high energy ($>$ 3 keV, in the
observer frame) excesses with respect to these models.  We estimated the
joint statistical significance of these excesses at the level
of 4.3 $\sigma$.  In all four cases, the deviations can be modeled
well by adding either a second power law or a blackbody component to
the usual synchrotron power law spectrum. The additional power law
would be explained by the emerging of the afterglow, while the
blackbody could be interpreted as the photospheric emission from
X--ray flares or as the shock breakout emission. In one case these
models leave a 2.2$\sigma$ excess which can be fit by a Gaussian line
at the energy the highly ionized Nickel recombination.}
{Although the data do not allow an unequivocal interpretation, the
importance of this analysis consists in the fact that we show that a
simple power law model or a Band model are insufficient to describe
the X-ray spectra of a small homogeneous sample of GRBs at the end
of their prompt phase.}
 \maketitle
\section{Introduction}
The X-ray telescope (XRT, \cite{Burrows05}) on board the Swift
satellite (\cite{Gehrels04}) allows to perform time resolved
spectroscopy of a large number of gamma ray burst (GRB) afterglows in
the 0.3-10 keV energy band.  The vast majority of the spectra can be
modeled by a single absorbed power law (SPL) model.  In most of the
afterglows a strong spectral evolution is observed in the early phases
with spectral indexes varying in the range 0.5-5 (\cite{Obrien06},\cite{Butler07},
\cite{Zhang07}).  These observations are consistent with the classical
fireball model which describes the burst and afterglow emission due to
synchrotron radiation.
\begin{figure*}
\begin{tabular}{cc}
\includegraphics[width=8.0cm]{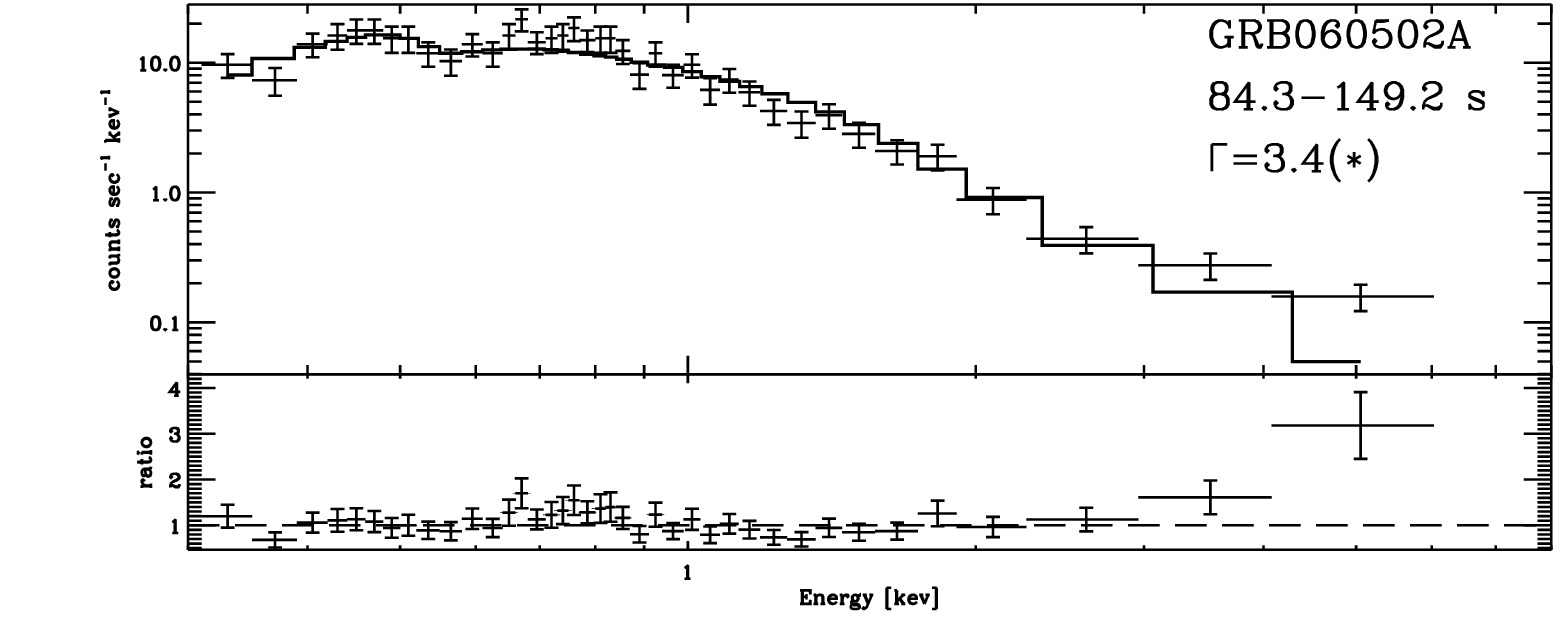}  & \includegraphics[width=8.0cm] {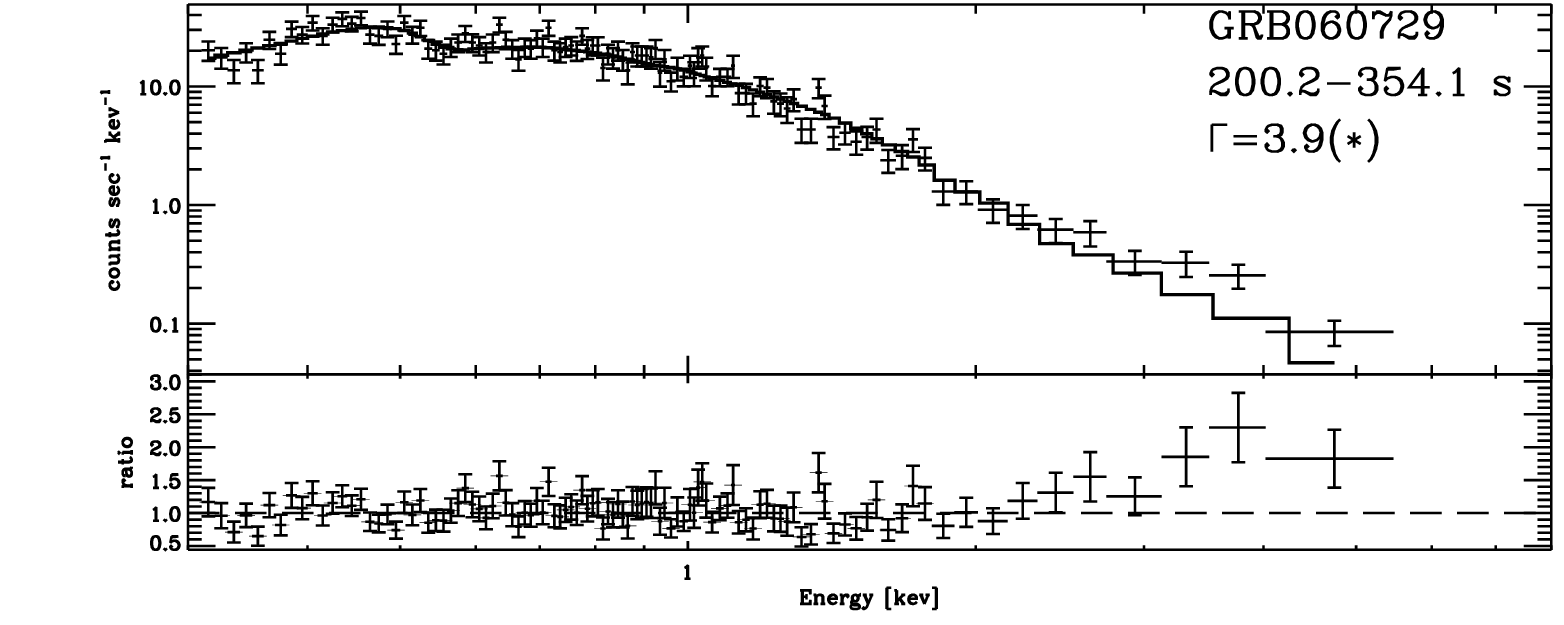}  \\ 
\includegraphics[width=8.0cm]{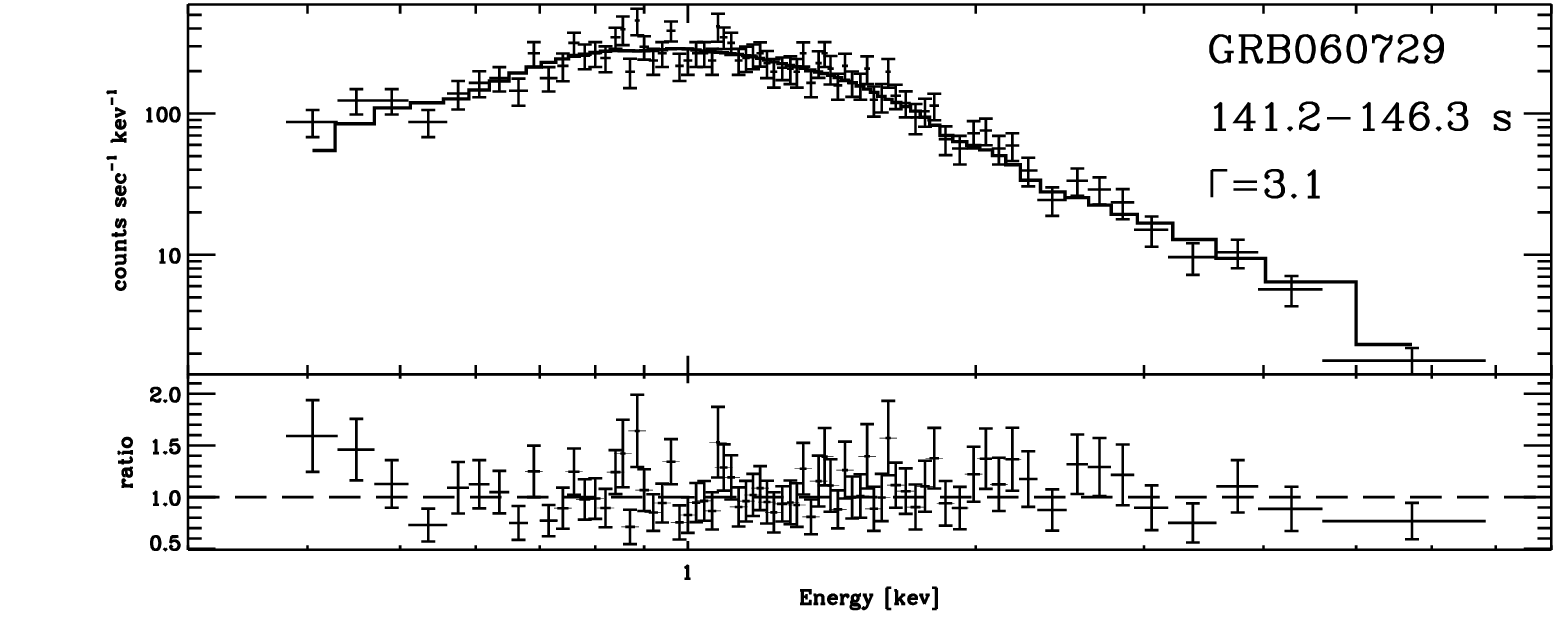}   & \includegraphics[width=8.0cm]{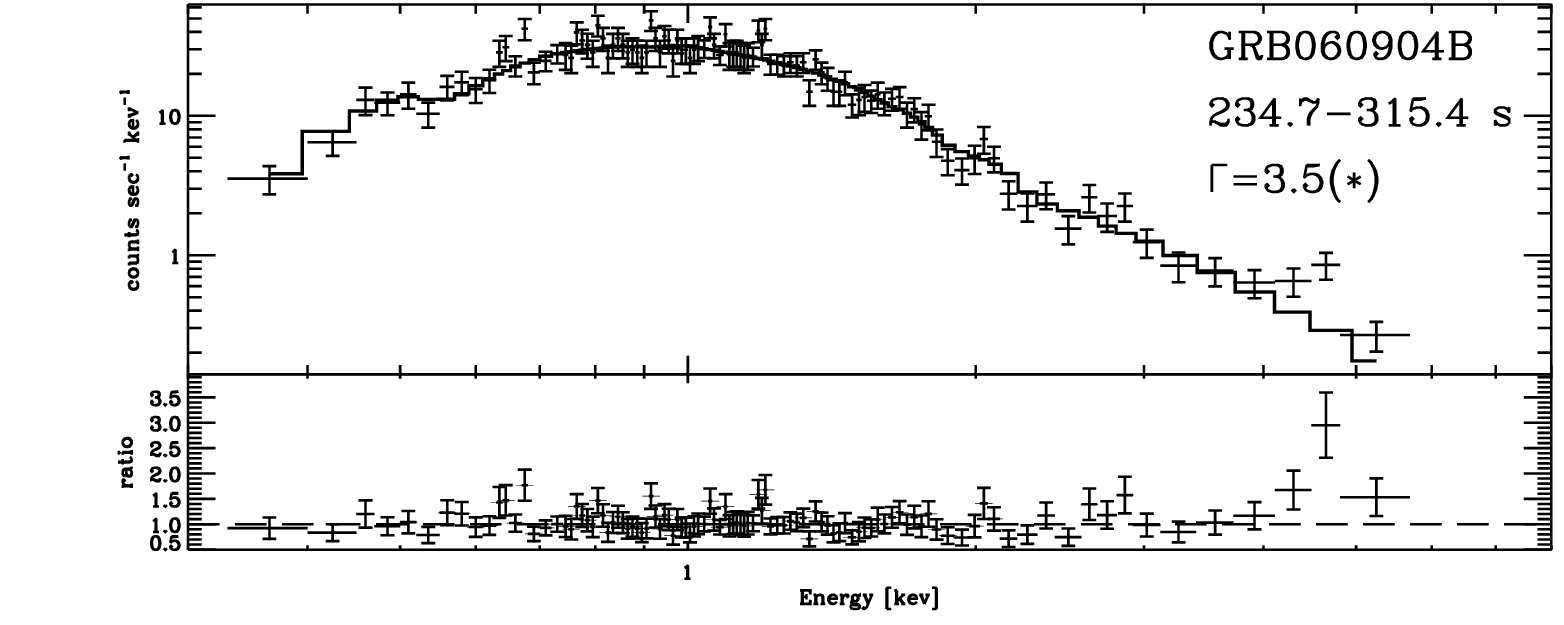}  \\  
\includegraphics[width=8.0cm]{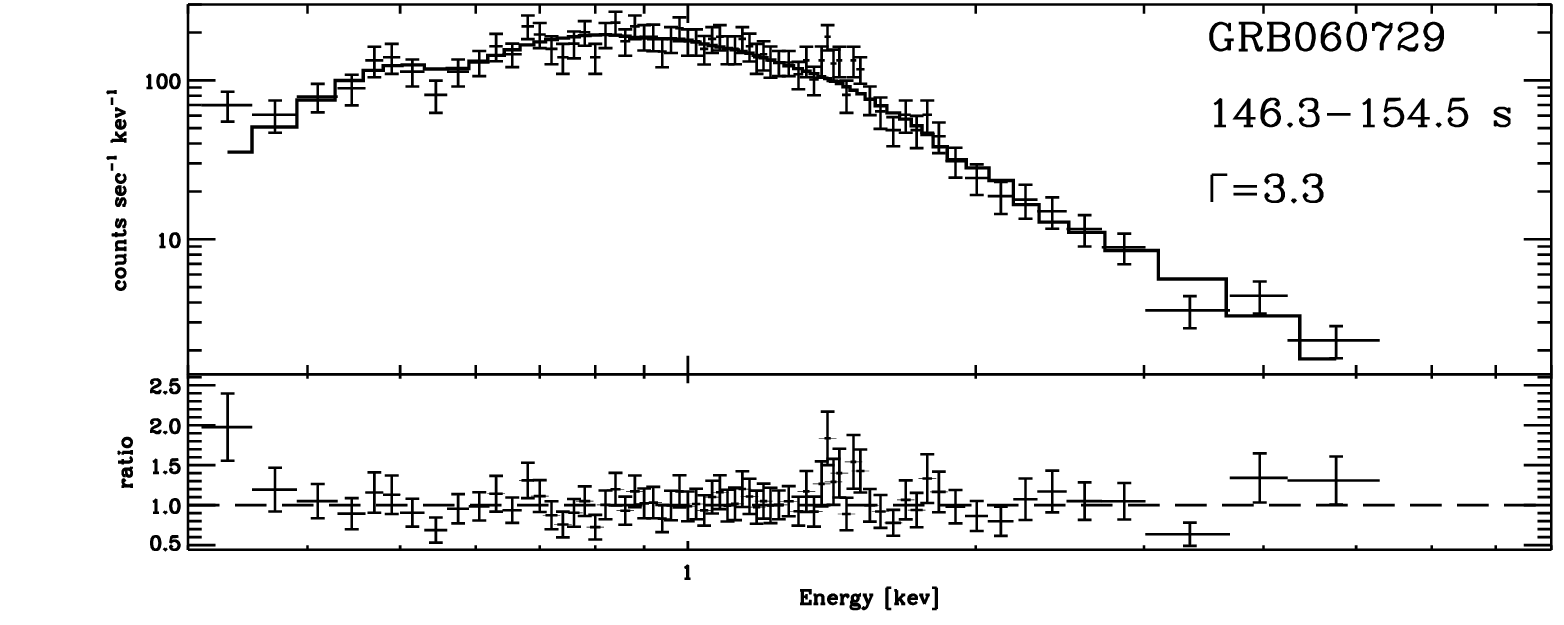}   & \includegraphics[width=8.0cm]{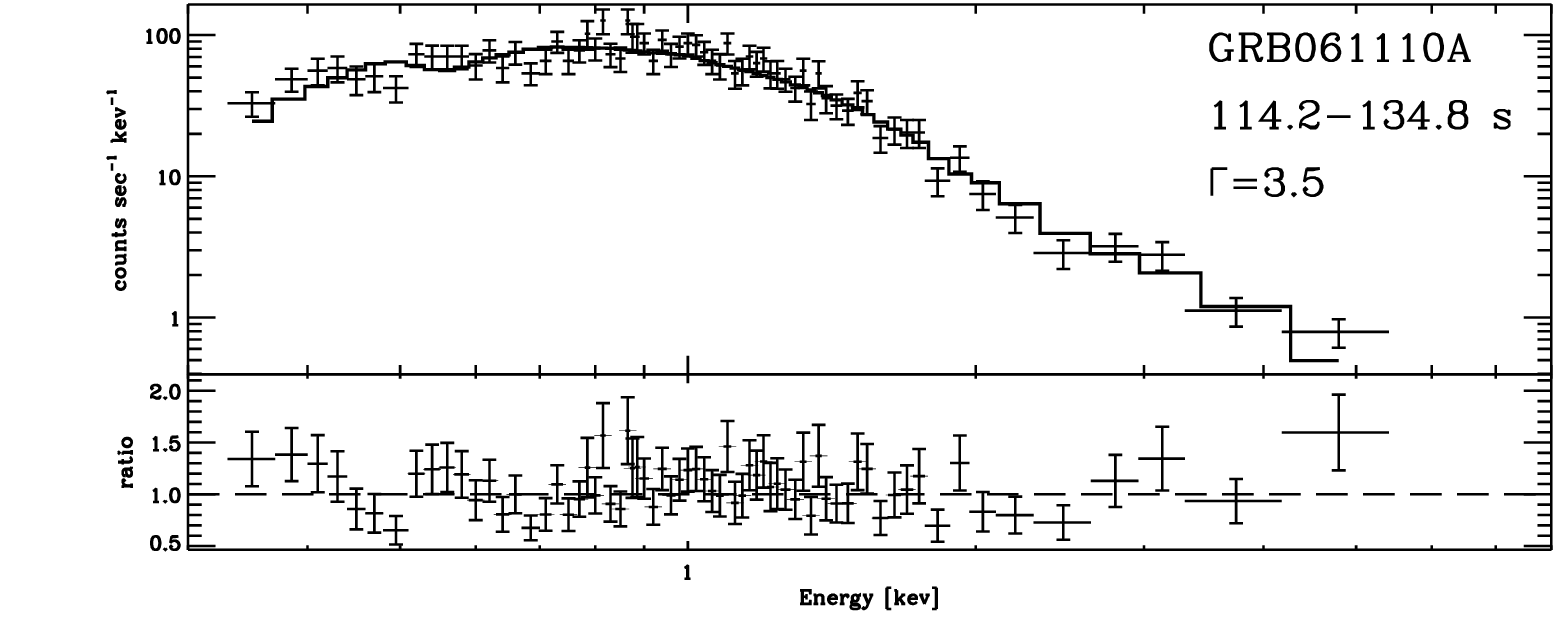}  \\  
\includegraphics[width=8.0cm]{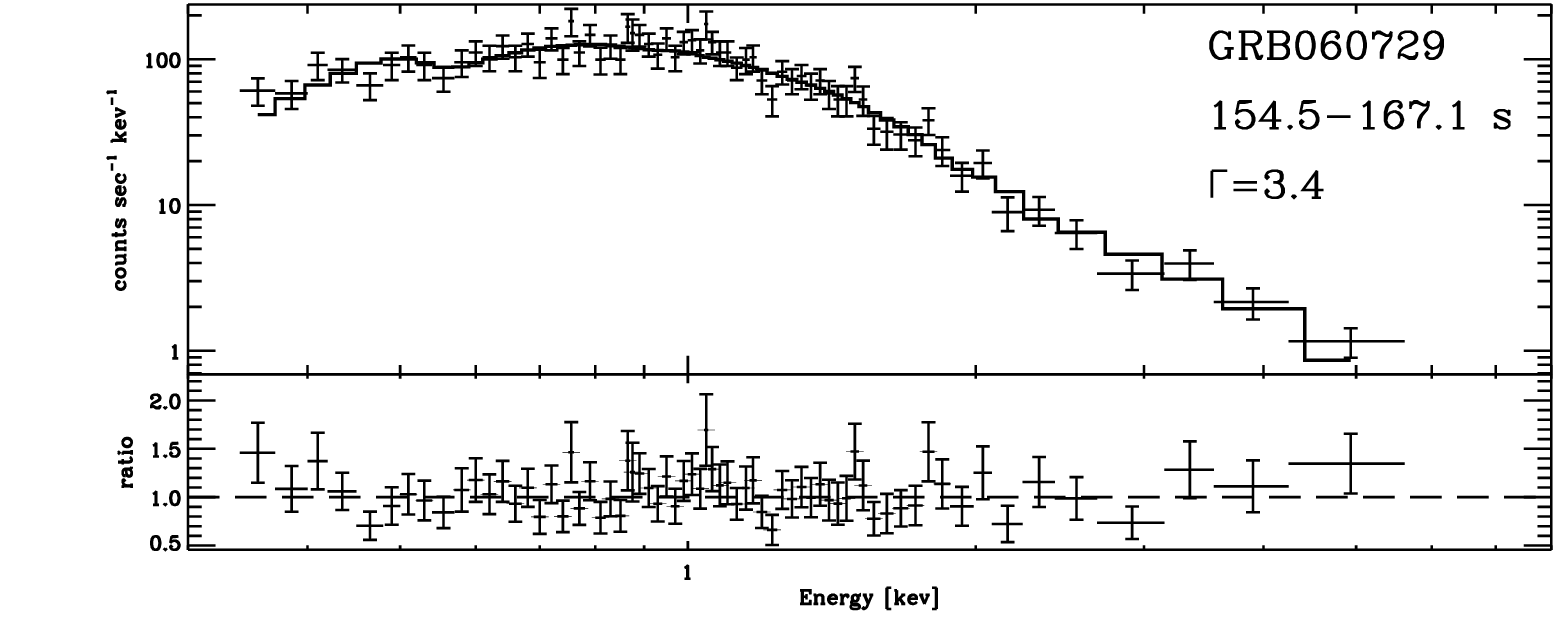}   & \includegraphics[width=8.0cm]{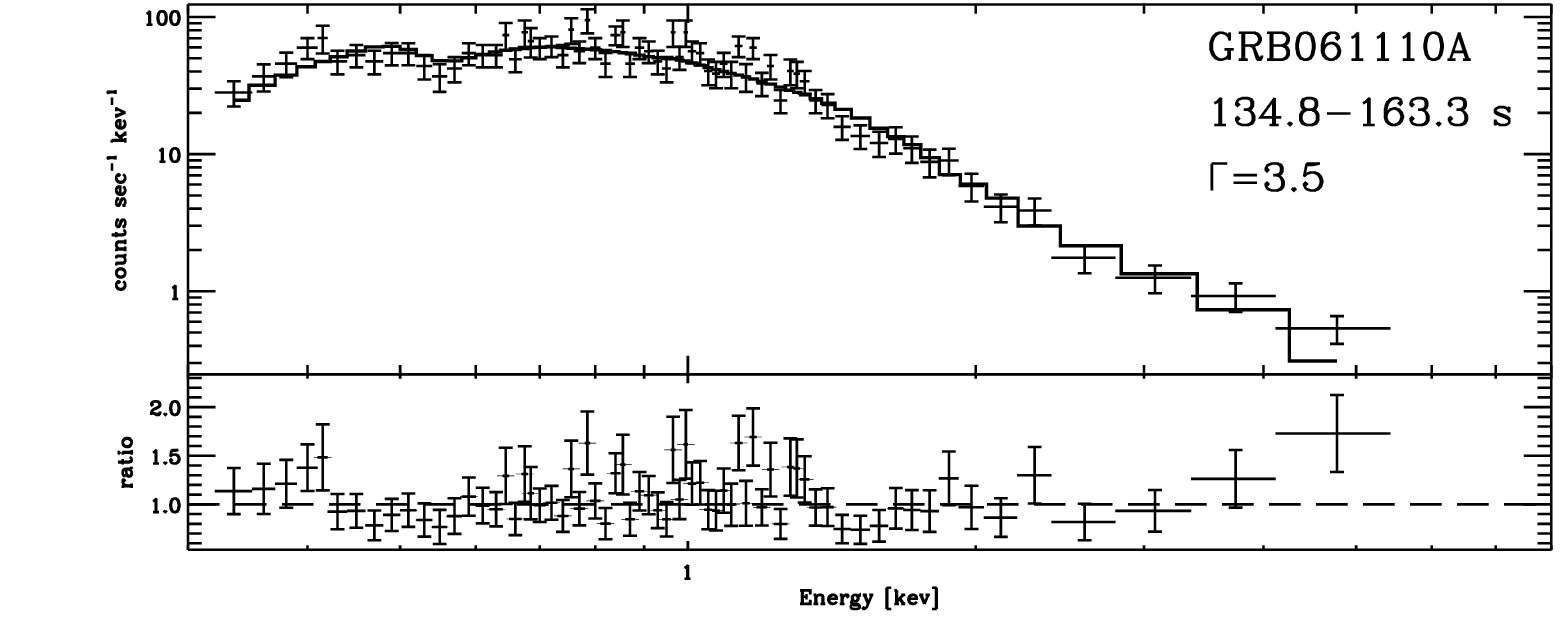}  \\  
\includegraphics[width=8.0cm]{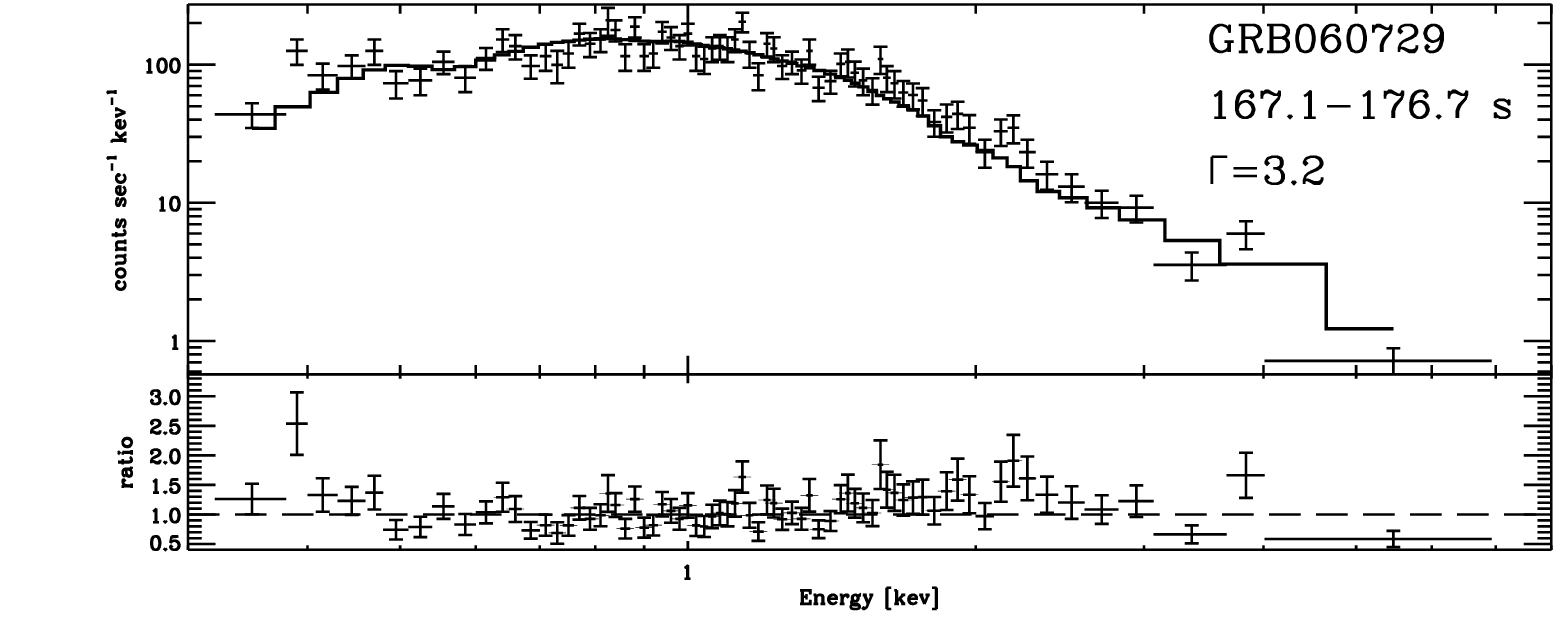}   & \includegraphics[width=8.0cm]{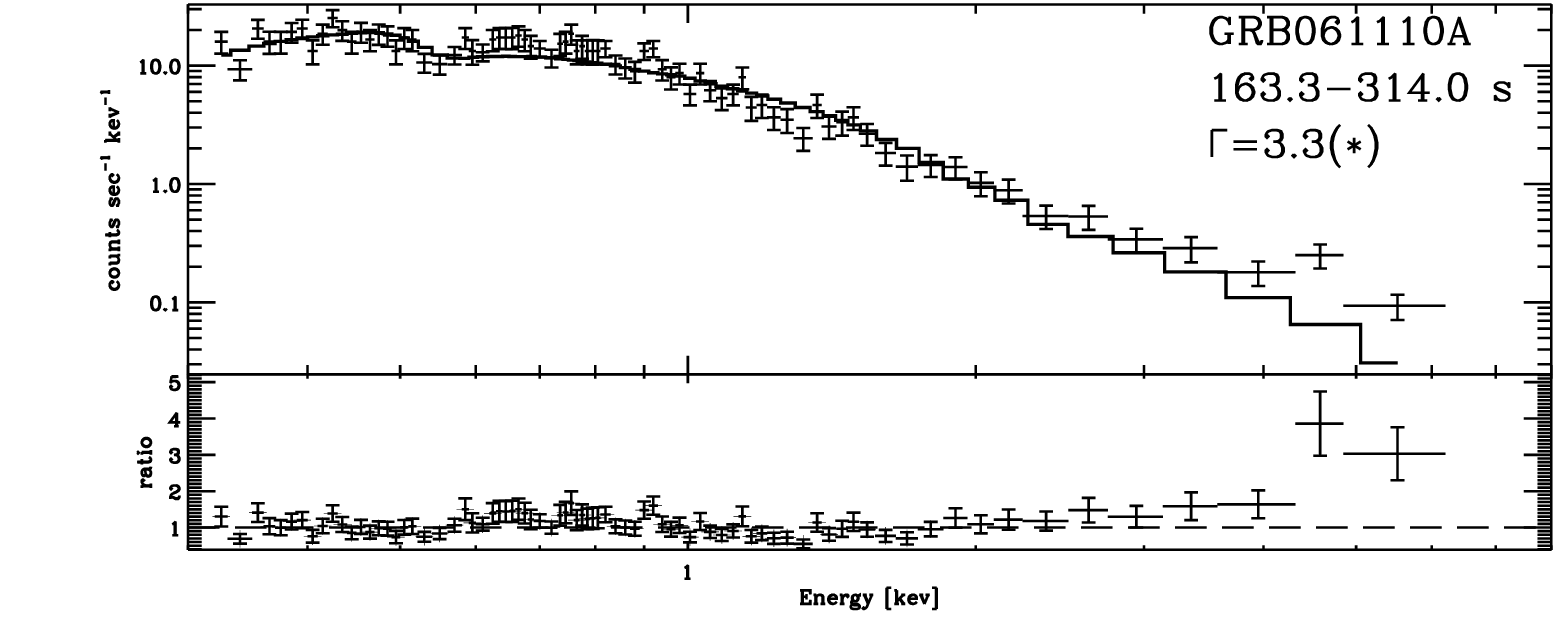}  \\  
\includegraphics[width=8.0cm]{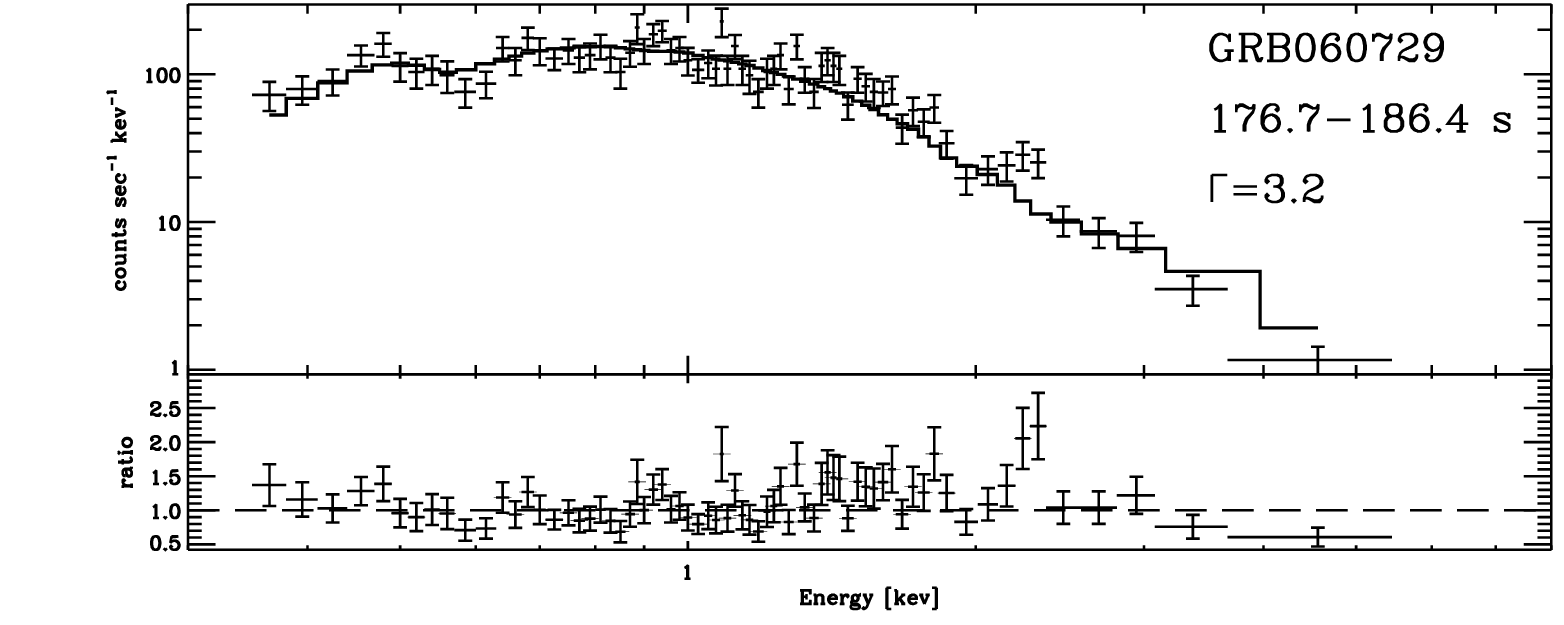}   & \includegraphics[width=8.0cm]{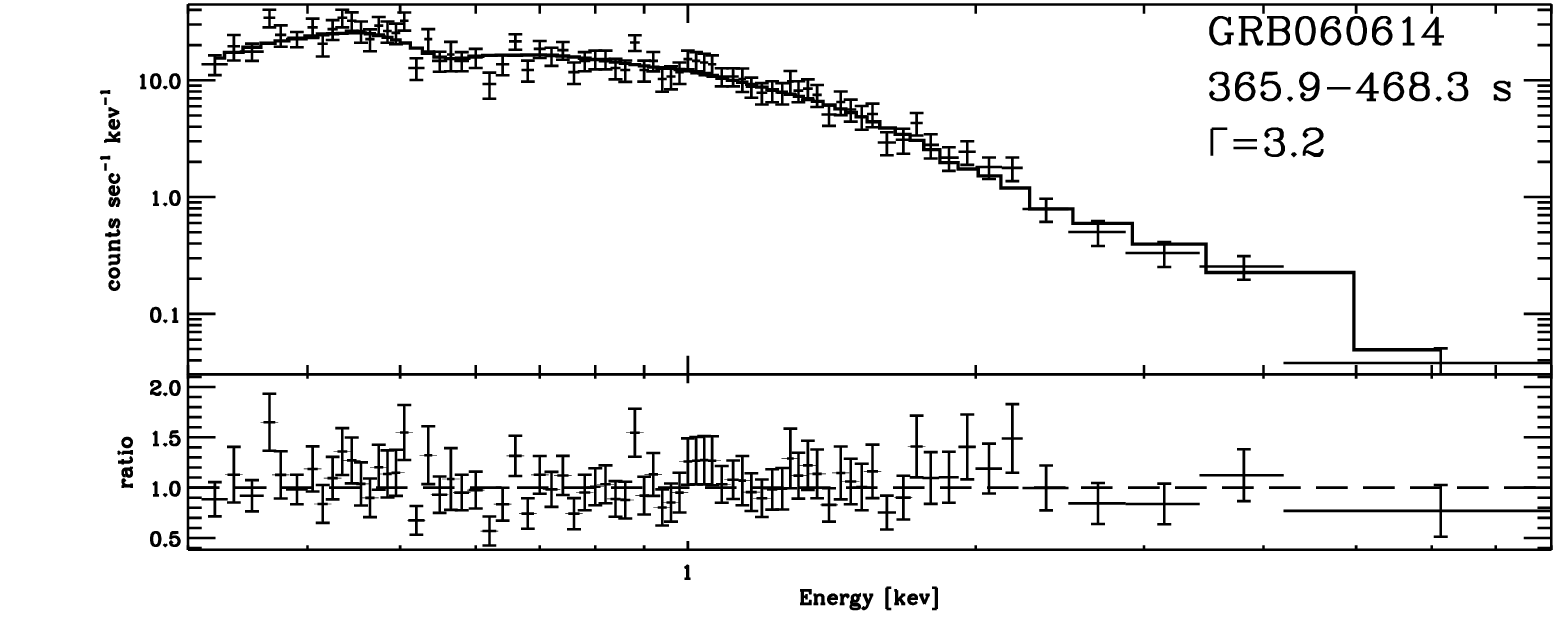}  \\  
\includegraphics[width=8.0cm]{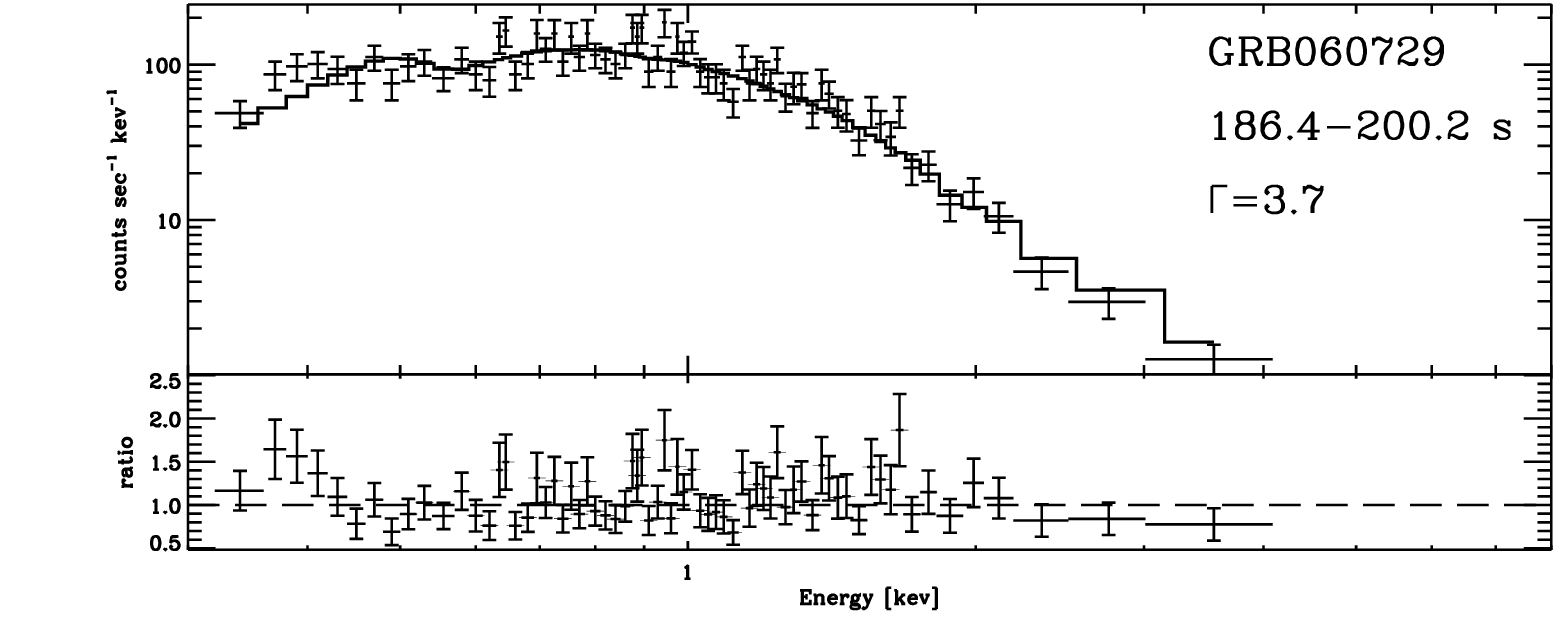}   &                                         \\  
\end{tabular}
\caption{The 13 very soft spectra are shown together with their best SPL fit. They are sorted in time from 
the top-left corner.  The observed time interval 
and the photon index are indicated in the figures. The asterisks denote the four spectra for which we
found significant departures from SPL model.}
\label{fig:f1}
\end{figure*}
It is possible that detection of different spectral features, signature of other
emission mechanisms, like thermal components or recombination lines
might give some insight about GRB progenitors, chemical composition,
physical conditions and geometry of the GRB environment.  
Before the Swift mission, several X--ray line detections were reported in GRB
late ($>$ 10 hr) afterglow observations. The statistical significance
of these detections has been questioned by \cite{Sako05}, who
concluded that there were no credible X--ray features in any GRB
afterglow. Butler et al. (2005), however, showed that for GRB011211 the 
different estimates of the statistical significance can be explained
by the different approach in the continuum modelling.

In the Swift afterglow observations only a few deviations from SPL
have been reported.  Most of them have been explained by the curvature
of the synchrotron spectrum and the presence of the $\nu F_{\nu}$ peak
(E$_{\rm{peak}}$) within the XRT band (Falcone et al. 2006, Butler \&
Kocevski 2007, Goad et al. 2007, Mangano et al. 2007, Godet 2007a).

Moreover Butler (2007) found anomalous soft X-ray emission in the
spectra of four GRBs which can be interpreted as multiple emission
lines due to K shell transition in light metals as well as thermal
emission from a blackbody with temperature $\sim$0.1 keV.
\cite{Grupe07} and Godet et al. (2007a) found that early afterglow data 
of GRB060729 and GRB050822, respectively, can be fitted with a SPL plus 
a blackbody with decreasing temperature in the first few hundreds 
seconds from the beginning of the prompt emission. Campana et al. (2006) 
found a cooling thermal component in the spectrum of SN2006aj/XRF060218,
interpreted as due to the supernova shock break-out (this interpretation
has been subsequently questioned by Li 2007 and Ghisellini et al. 2007).

Starting from the idea that any deviation from a SPL spectral model,
if present, would be a faint signal mostly covered by the high level
``noise'' of synchrotron emission, we searched for the best
observational conditions to detect it. In particular, we searched for
high energy excesses with respect to the SPL when the spectrum is
steepest and, at least in the hard part of the energy band, could be
the non dominating component.

Throughout this paper, all errors are quoted at 68\% confidence level
for one parameter of interest, unless otherwise specified. The reduced
$\chi^2$ will be denoted as $\chi^2_\nu$ and the number of degrees of
freedom with the abbreviation ``d.o.f.''.  We follow the convention
$F_\nu(\nu,t) \propto t^{-\alpha}\nu^{-\beta}$, where $\alpha$ and
$\beta$ are the temporal decay slope and the spectral index,
respectively. As time origin, we will adopt the Burst Alert Telescope
(BAT) trigger (T0).  The photon index is $\Gamma = 1 + \beta$.  Last,
we adopt the standard ``concordance'' cosmology parameters,
$\Omega_{\rm m} = 0.27$, $\Omega_\Lambda = 0.73$, $h_0 = 0.71$.
\section{Data analysis and sample selection} 
We considered the GRB sample detected by BAT, promptly observed by
XRT in the period April 2005--January 2007 which collected at least 800 photons. 
We restricted our analysis to the sample with spectroscopically measured
redshift in order to separate the local contribution
to the total absorbing column from the Galactic one.
We excluded GRB060218 from our analysis because of its peculiarity
(\cite{Campana06}). For each burst, when possible, we split the XRT
data in different time intervals in such a way that in each of them
there are 2000 photons, before the background subtraction (the last
spectrum collects the remaining photons).  The final sample
consists of 22 bursts for a total of 214 time intervals.  The data
reduction was performed using the standard software (HEADAS software,
v6.1, CALDB version Jul07) and following the procedures reported in
the instrument user guides\footnote{http://heasarc.nasa.gov/docs/swift/analysis/\#documentation}.
The spectral analysis was performed using \texttt{XSPEC} (v11.3).  The
214 spectra were binned in order to ensure a minimum of 20 counts per
energy bin, ignoring channels below 0.3~keV and above 10~keV.  For the
Galactic hydrogen column density $N_{\rm H,MW}$ we assumed the value
reported in \cite{DickeyLockman90} along the GRB direction. We multiplied
to the spectral models an extra neutral absorber at the source
redshift letting the column density $N_{{\rm H},z}$ free to vary.
SPL provided very good fits in most cases and we found that the photon
index $\Gamma$ ranges from 0.5 to 3.9.  In particular there are 13
spectra (6\% of the total), belonging to 5 different bursts with
$\Gamma > $3 .  These are: one from GRB 060502A, GRB 060614 and GRB
060904B, three from GRB 061110A and seven from GRB 060729.  
They are all collected in Windowed Timing (WT)
mode (see Hill et al. 2004 for a description of the different
operational modes of the XRT) and belong to the prompt--afterglow
transition phase, observed less than 500 seconds after the event
triggers. In Fig.~\ref{fig:f1} the 13 very soft spectra are shown
together with their best SPL fit.  The mean $\chi^2_\nu$ value of this
13 spectra sub-sample is 1.20$\pm$0.07, significantly worse than the
average of the entire sample that is 1.01$\pm$0.01 (chance probability
$<$10$^{-4}$). In particular there are some evident departures from SPL
model at high energies in four spectra from four bursts.
\section{Excess statistical significance \label{sect:s3}}
\begin{figure}
\includegraphics[width=9cm]{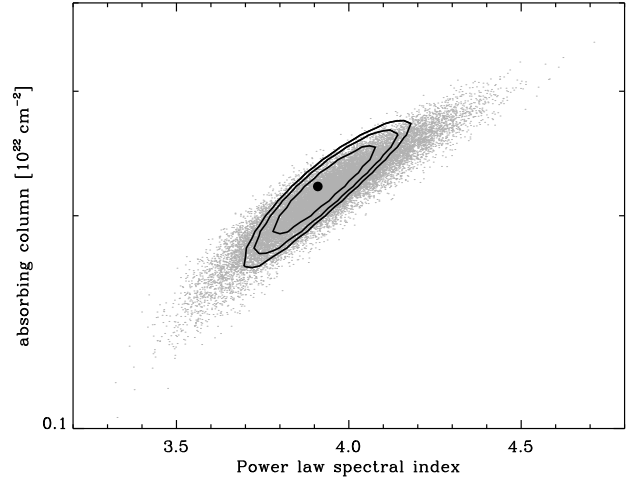}
\caption{
Black point and black contour are the best fit and 1,2,3 $\sigma$ confidence
contour of the fit on one observed spectrum of GRB 060729. Grey dots are
the best fit results on a sample of 20,000 simulated spectra.
}
\label{fig:f2}
\end{figure}
Having in mind the {\it rules of thumb} given by Protassov et
al. (2002), to determine if there are statistically significant
departures from SPL, we followed step by step the method described by
\cite{Rutledge02} which was used to calculate the significance of many
X-ray features by \cite{Sako05}.  
\begin{figure*}
\begin{tabular}{cc}
\includegraphics[width=9cm]{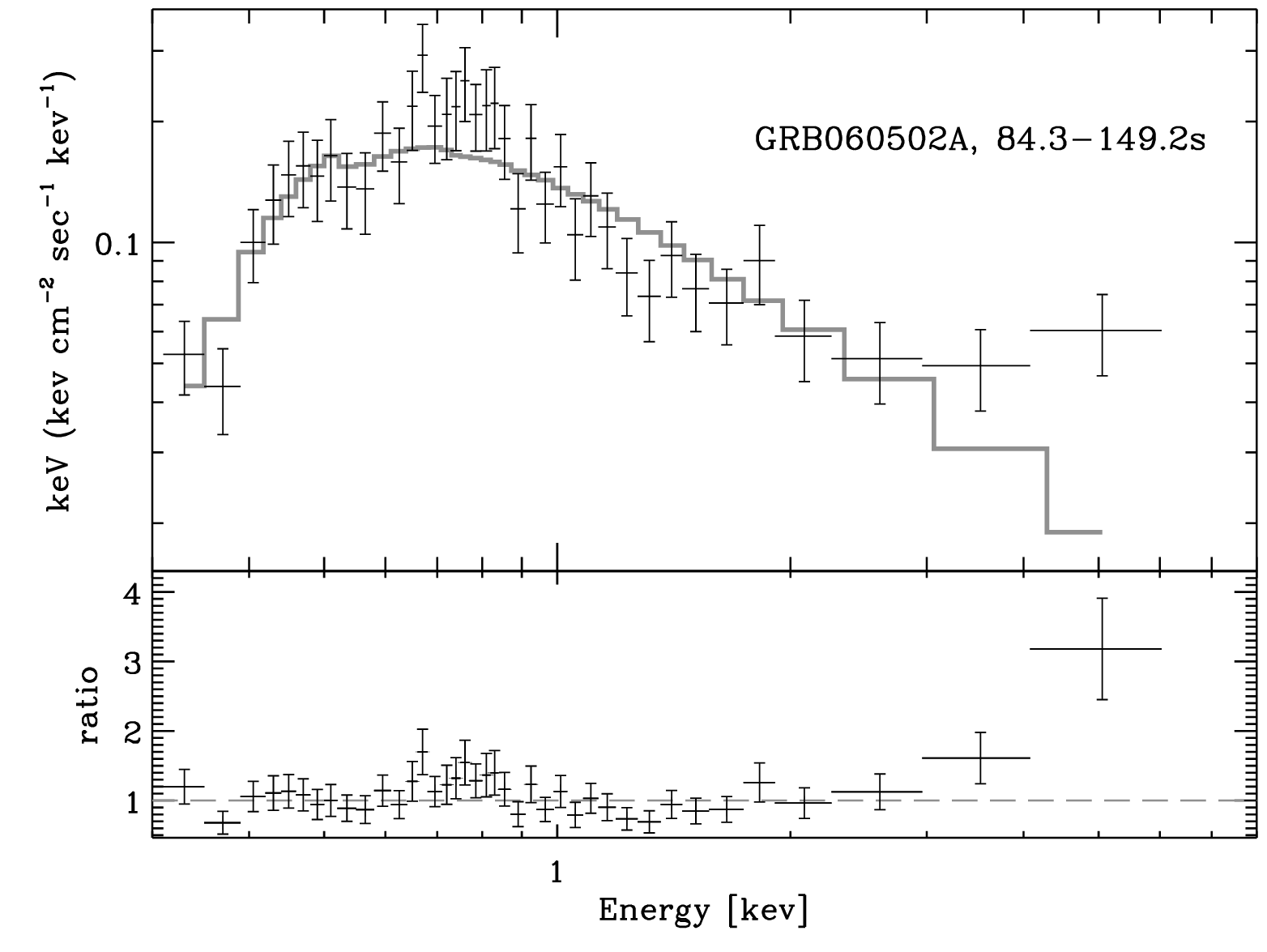} & \includegraphics[width=9cm]{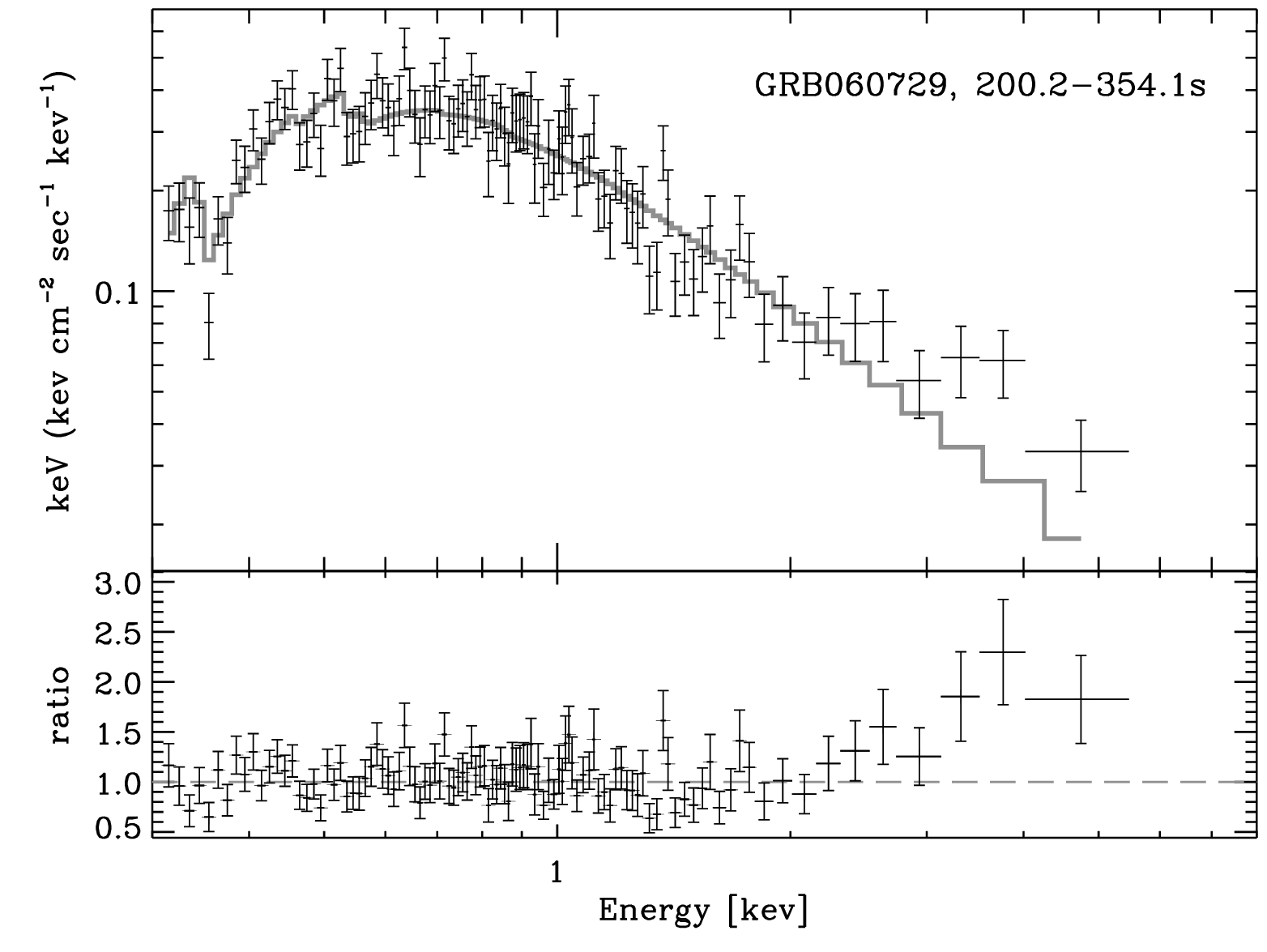} \\
\includegraphics[width=9cm]{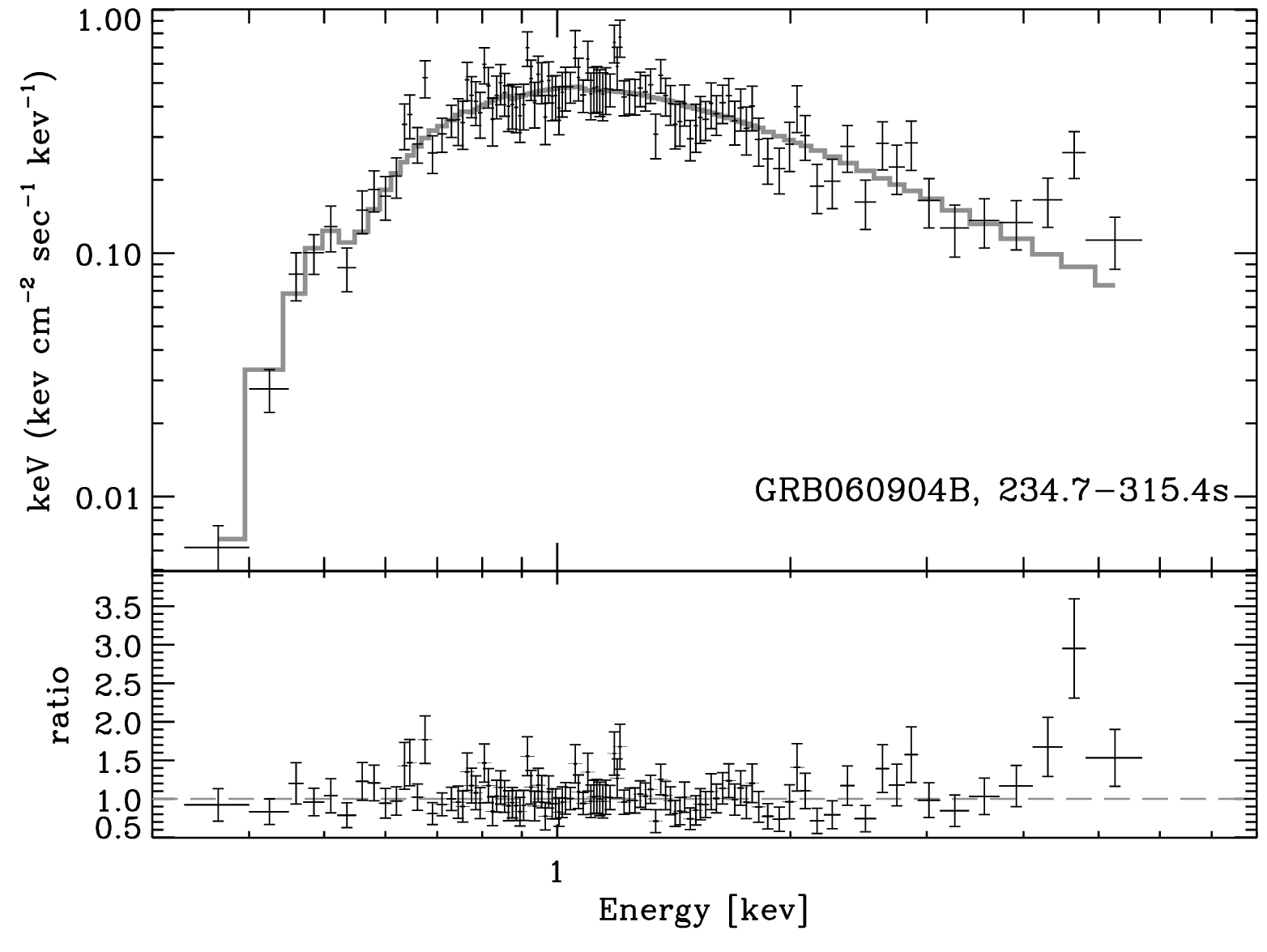} & \includegraphics[width=9cm]{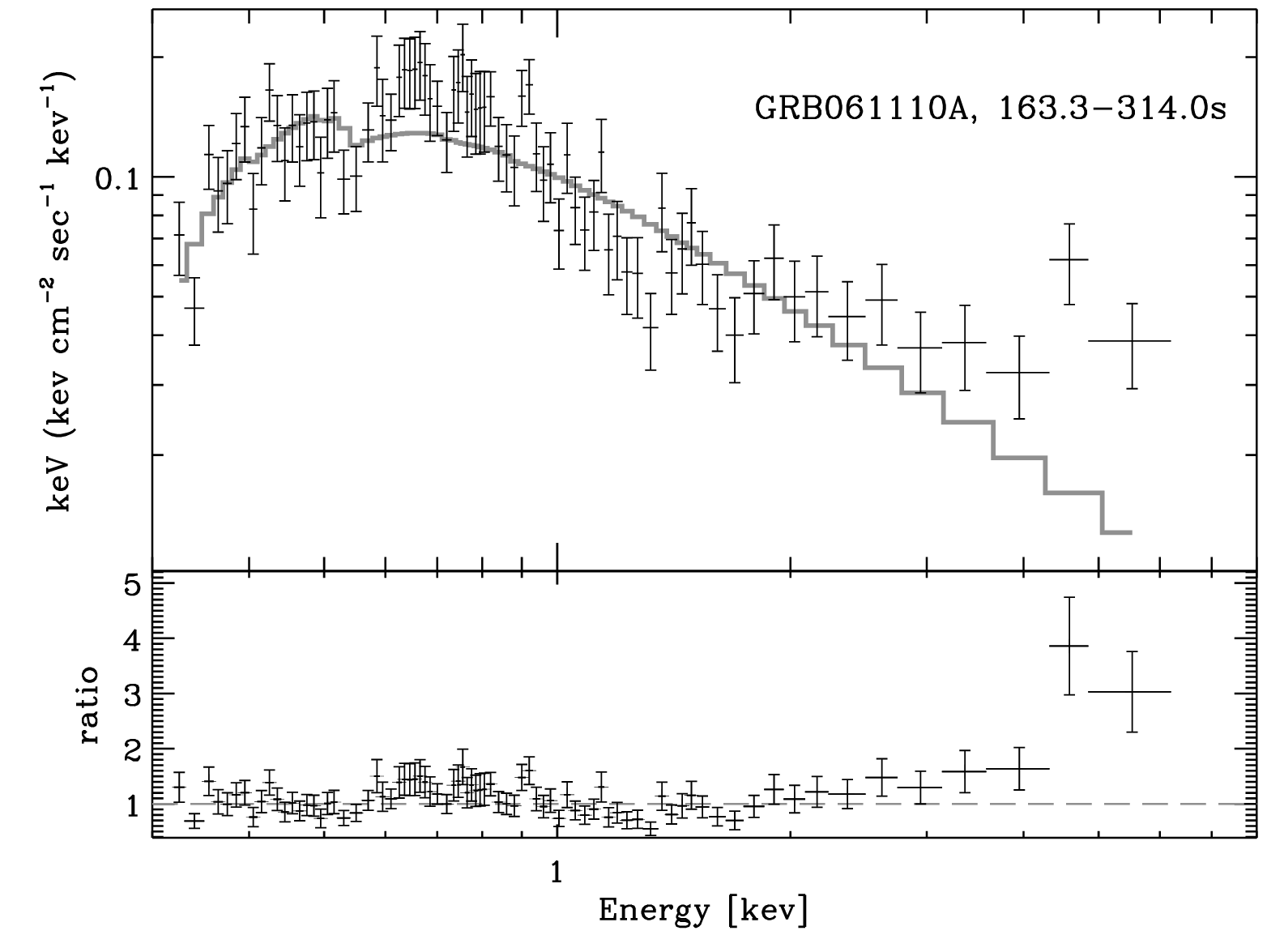} \\
\end{tabular}
\caption{ The spectra (in energy units) of the four GRB afterglows for which we detect
significant departures from SPL model which is plotted in grey.}
\label{fig:f3}
\end{figure*}
First, we considered the redistribution matrix (RMF=RMF(PI,E)) and we
fitted each column of the matrix (2400 in total;
see\footnote{http://heasarc.gsfc.nasa.gov/docs/heasarc/caldb/swift/})
by a Gaussian function. As done by \cite{Rutledge02} we built a new
RMF with the 2400 columns replaced by the Gaussian fit of the original
RMF.  Because our features are broader than the instrumental spectral
resolution (which is 0.11 keV at 4 keV) we also built four artificial
RMFs with Gaussian functions 3,5,10,16 times wider than the best fit of
the original RMF value. In practice, we built a set of 4 different RMF
worsening the spectral resolution to look for the scale which
maximizes the signal-to-noise ratio.

For each of the 13 spectra being tested, we convolved the PI count
spectrum with the 5 (1 nominal, 4 smoothed) RMFs.  Then, for each of
the 13 observed spectra, we created 100,000 Montecarlo (MC)
realizations of the raw pulse-invariant (PI) spectra based on the
respective best-fit SPL models using a fixed number of source plus
background photons. Background events were randomly selected from a
background spectral model, derived from fits to spectra obtained from
a source--free region of a deep exposure.
\begin{figure*}
\begin{tabular}{cc}
\includegraphics[width=9cm]{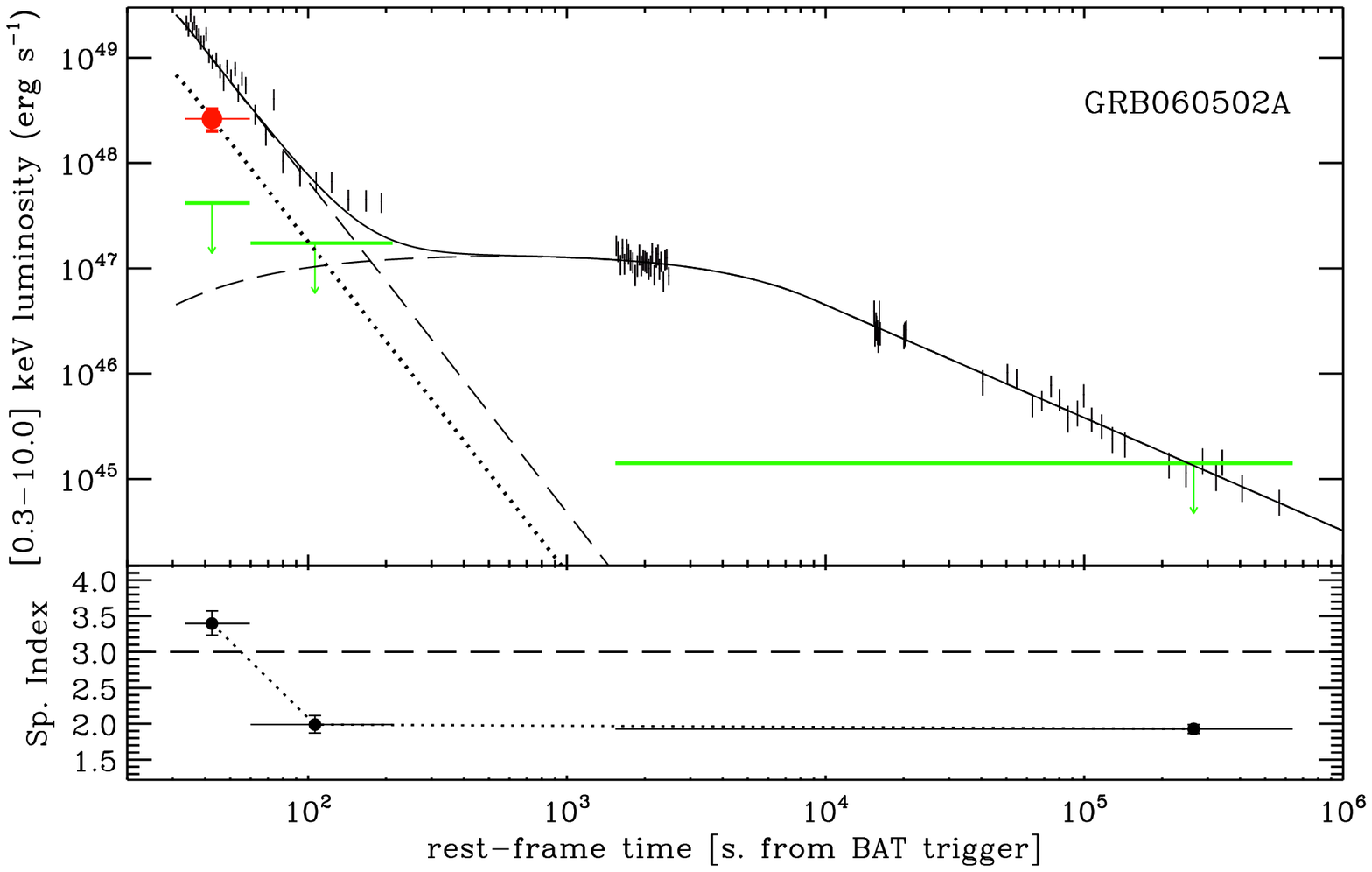} & \includegraphics[width=9cm]{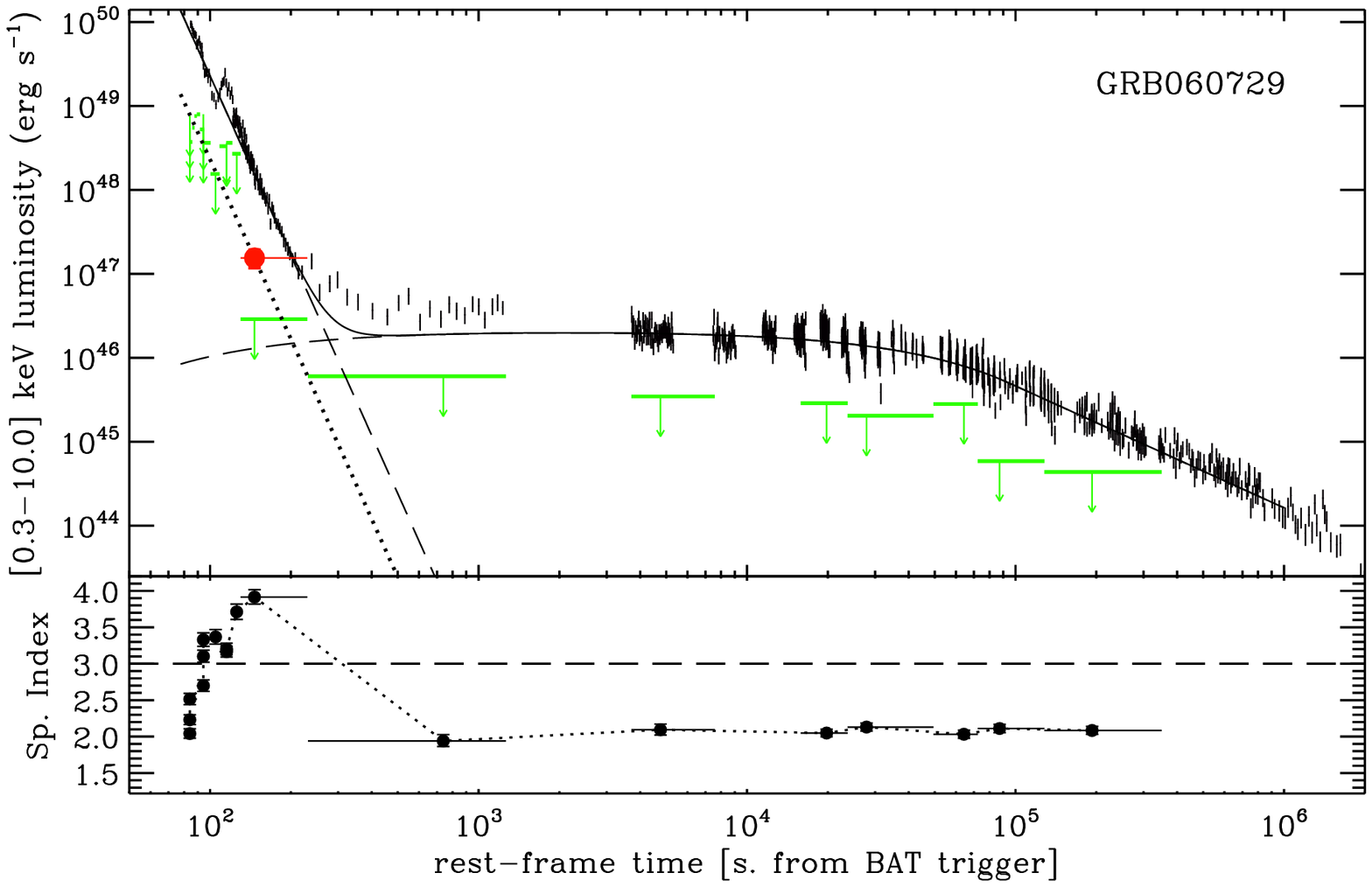} \\
\includegraphics[width=9cm]{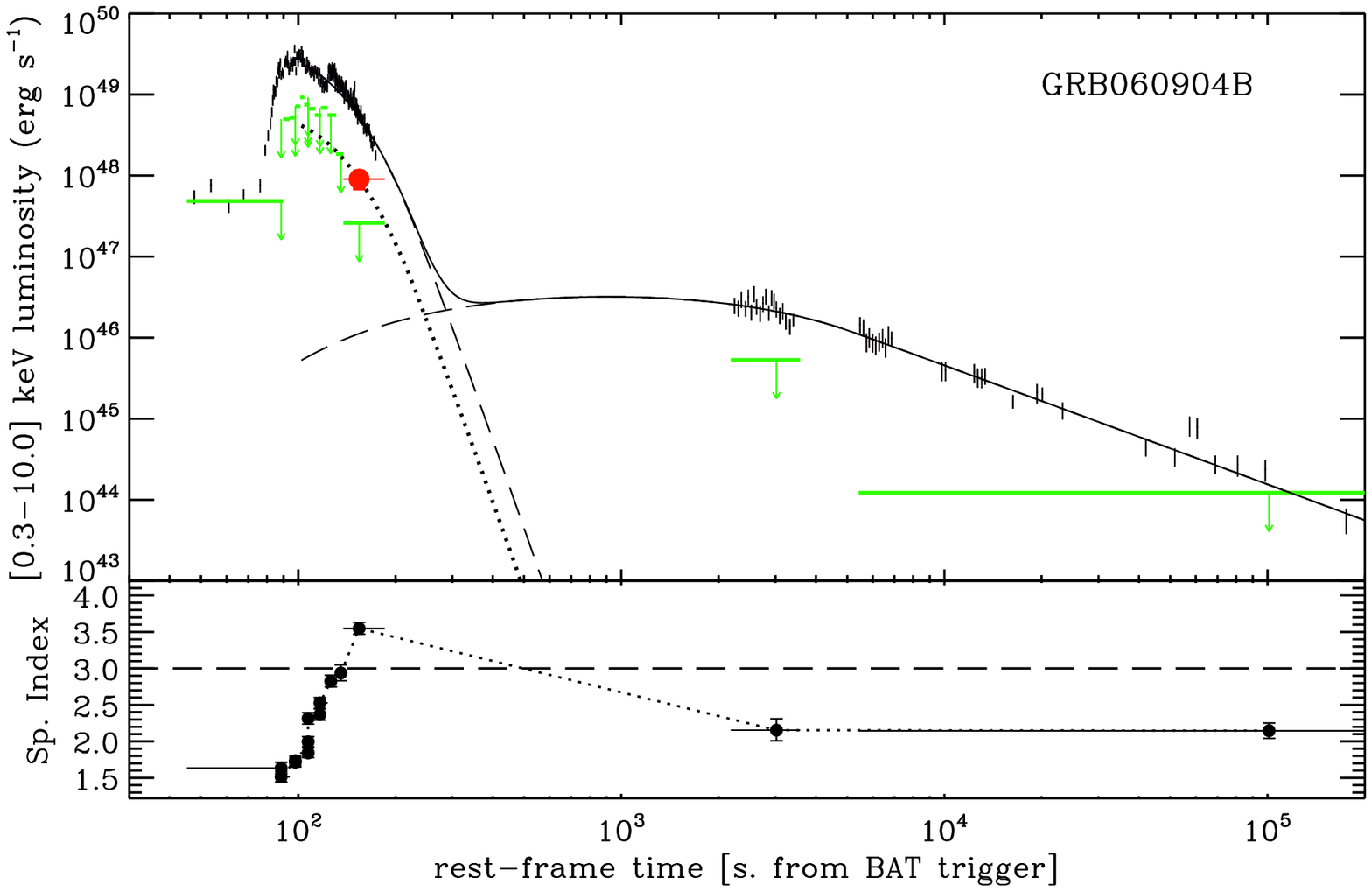} & \includegraphics[width=9cm]{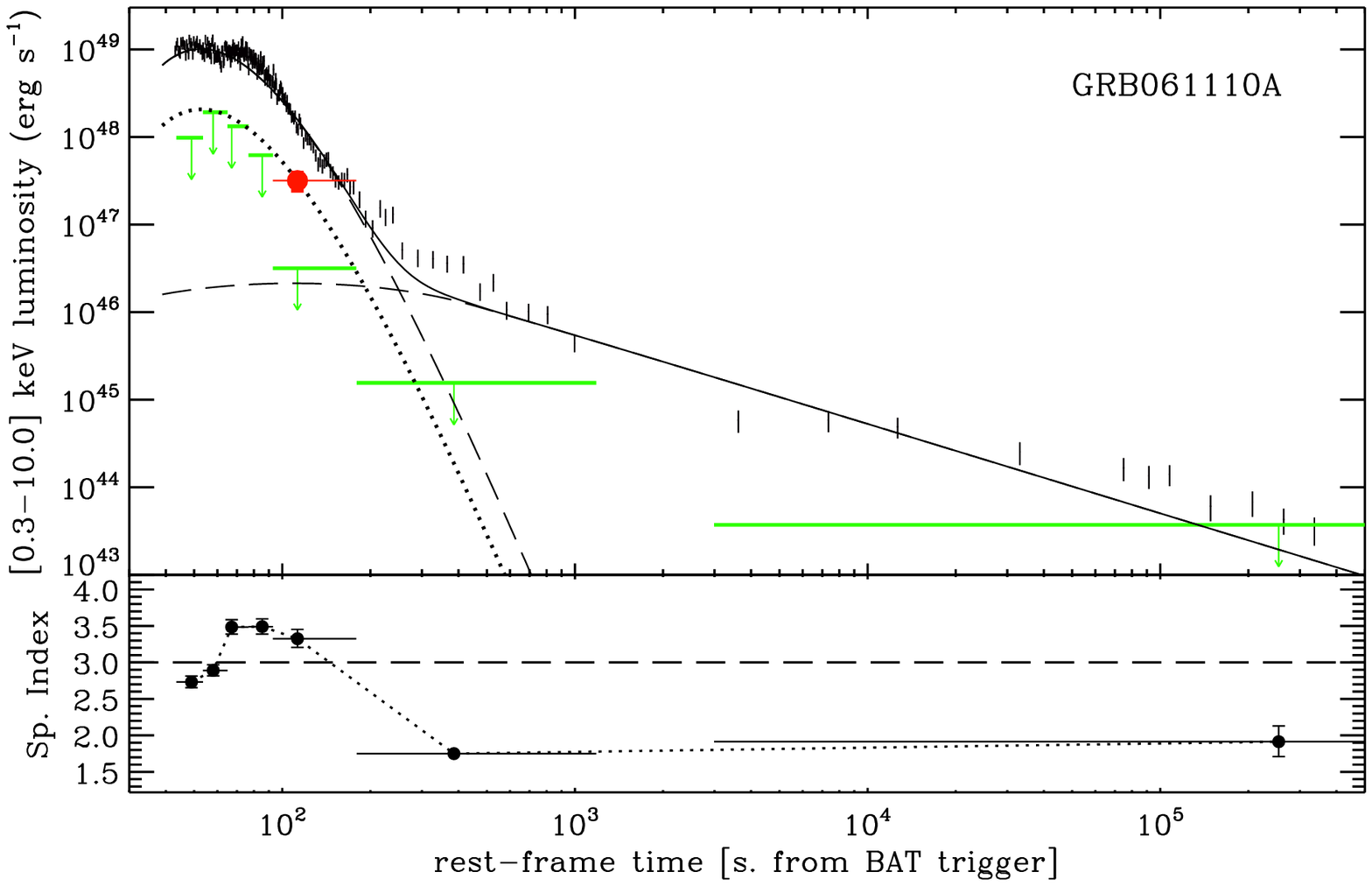} \\
\end{tabular}
\caption{ {\bf Upper panels:} The luminosity (0.3-10 keV rest--frame
energy band) curves of the four GRB afterglows for which we detect
significant departures from SPL model. With the continuous lines we
plot the best fit using the model described by \cite{Willingale07},
split in prompt and afterglow contributions (dashed lines).  With the
filled circles we plot the luminosity of the spectral excesses we
detect.  The upper limits to the detection of the extra components
(see Section \ref{sect:extra}) in the different time slices are also
reported as top-down arrows (for the clarity of the picture we report
only the value calculated assuming a DPL model, see Section
\ref{sect:extra}).  With the dotted line we plot the fit of the prompt
emission re-normalized to the luminosity of the extra component.  {\bf
Lower panels:} The spectral indexes $\Gamma$ given by SPL fit,
measured in the different time slices.}
\label{fig:f4}
\end{figure*}

To check the accuracy of our MC simulations for each GRB we ran the
XSPEC grouping and fit procedures on a sample of 20,000 simulated
spectra.  In Fig.~\ref{fig:f2} for one spectrum of GRB060729 we
compare the results of the SPL model fit of the observed data, with
the results of the fit on simulated data.

As we did for the observed data, we convolved all the 100,000
simulated PI spectra with the 5 matrices corresponding to the 5 filter
scales.  For each spectrum and each scale we counted the number of MC
realizations exceeding the feature in the data and recording the
energy and the scale for which this number is minimum.  There are four
spectra, in the sample of 13, for which we found excesses in the data
that can be reproduced as statistical fluctuations of the SPL in less
than 10 trials out of 100,000.  This correspond to a single trial
significance higher than 99.99\% . These are the last WT spectra,
coinciding with the last phase of the X--ray light curve steep decay,
of GRB 060502A, 060729, 060904B and 061110A (Table~\ref{tab:tab1} and
Fig.~\ref{fig:f3}, \ref{fig:f4}).  In the rest of the paper we will limit our
analysis to these four spectra. 
\begin{table}
\caption[]{ 
(I)  GRB: Name of the GRB;
(II) Rest--frame time interval;
(III) Redshift;
(IV) energy of the excess maximum (observer frame); 
(V) energy scale in times of instrumental resolution 
(VI)  Single trial significance;
(VII) Multi trial significance.
Ref: $^{1}${\cite {Laparola06}}; $^{2}${\cite {Cucchiara06}}; $^{3}${\cite {Grupe07}}; $^{4}${\cite {Thoene06a}};  
$^{5}${\cite {Grupe06}}; $^{6}${\cite {Fugazza06}}; $^{7}${\cite{Fox06}}; $^{8}${\cite {Thoene06b}}.}
\begin{tabular}{l|cccccc} 
\hline
\hline
GRB&$\Delta$t&z&E&Sc.&s.t.&m.t.\\   
&s&&keV&&(\%)&(\%)\\
\hline
060502A$^{1}$ & 33-59  &1.51$^{2}$ &5.7&7 &99.9932&99.4778\\                            
060729$^{3}$  & 130-230&0.54$^{4}$ &3.7&10&99.9875&99.0421\\
060904B$^{5}$ & 138-185&0.703$^{6}$&4.6&3 &99.9993&99.9461\\ 
061110A$^{7}$ & 93-179&0.757$^{8}$&4.7&5 &99.9997&99.9769\\
\hline
\end{tabular}
\label{tab:tab1}  
\end{table}

\begin{figure*}
\begin{tabular}{cc}
\includegraphics[width=9cm]{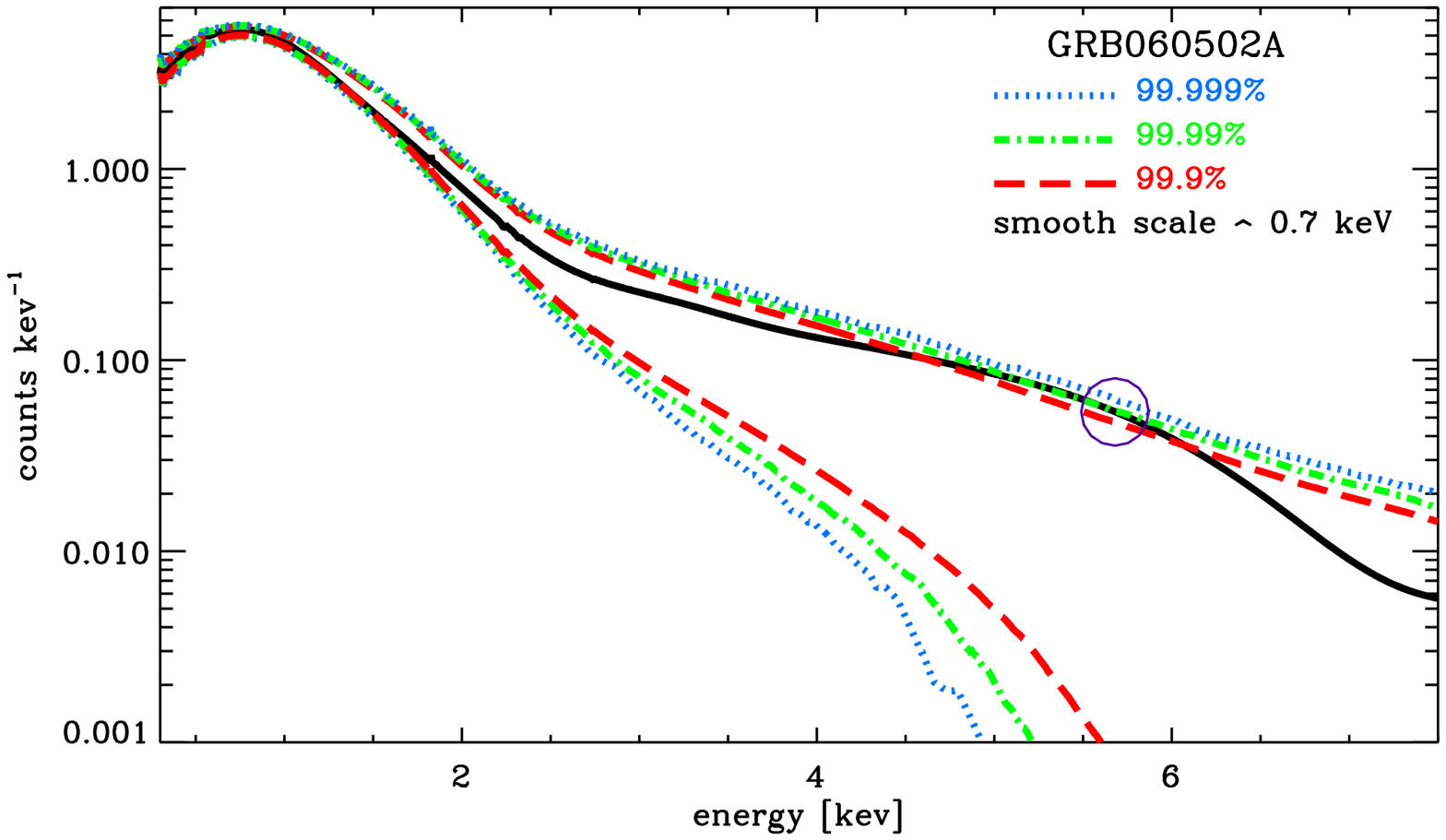} & \includegraphics[width=9cm]{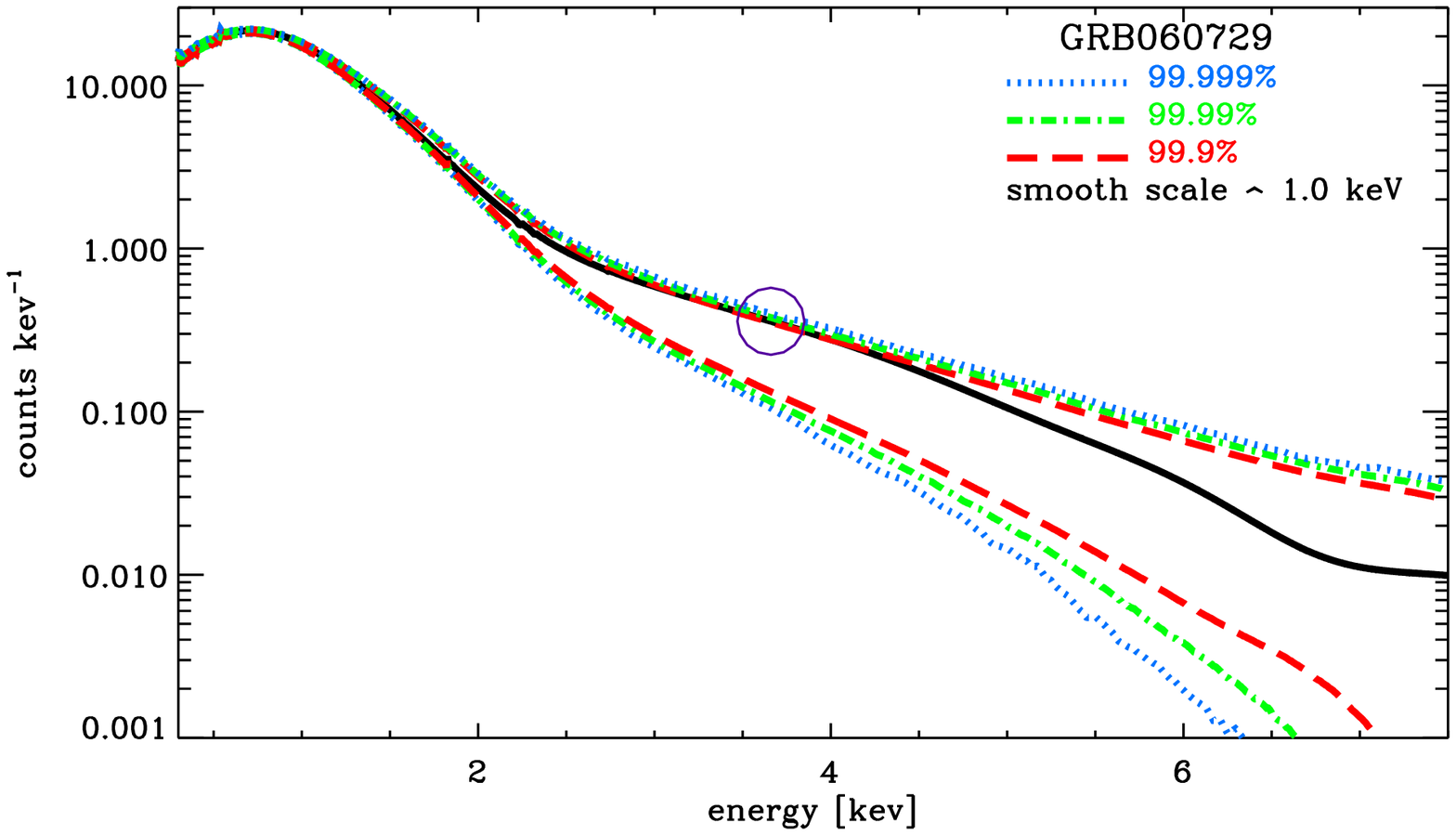} \\ 
\includegraphics[width=9cm]{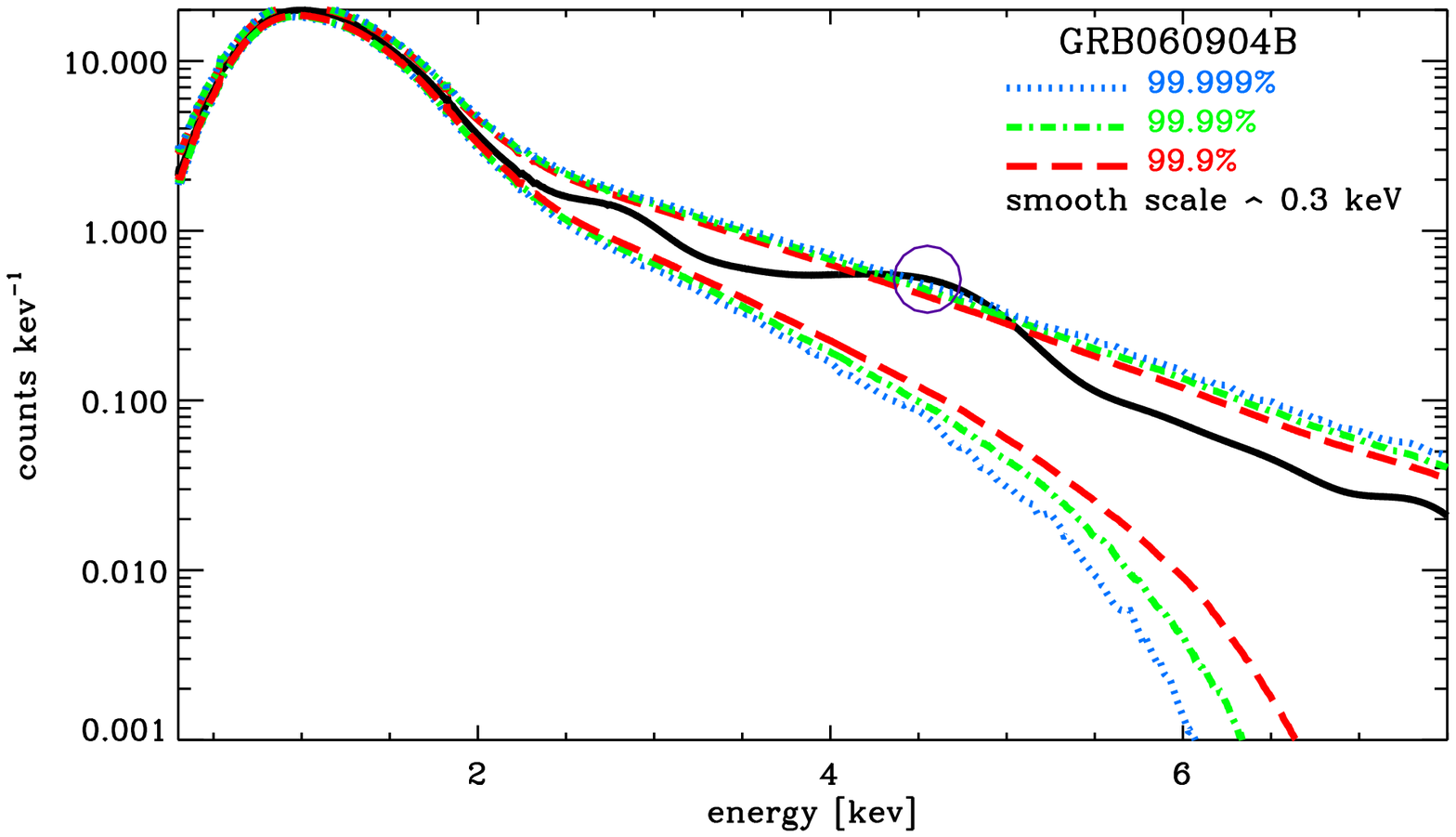} & \includegraphics[width=9cm]{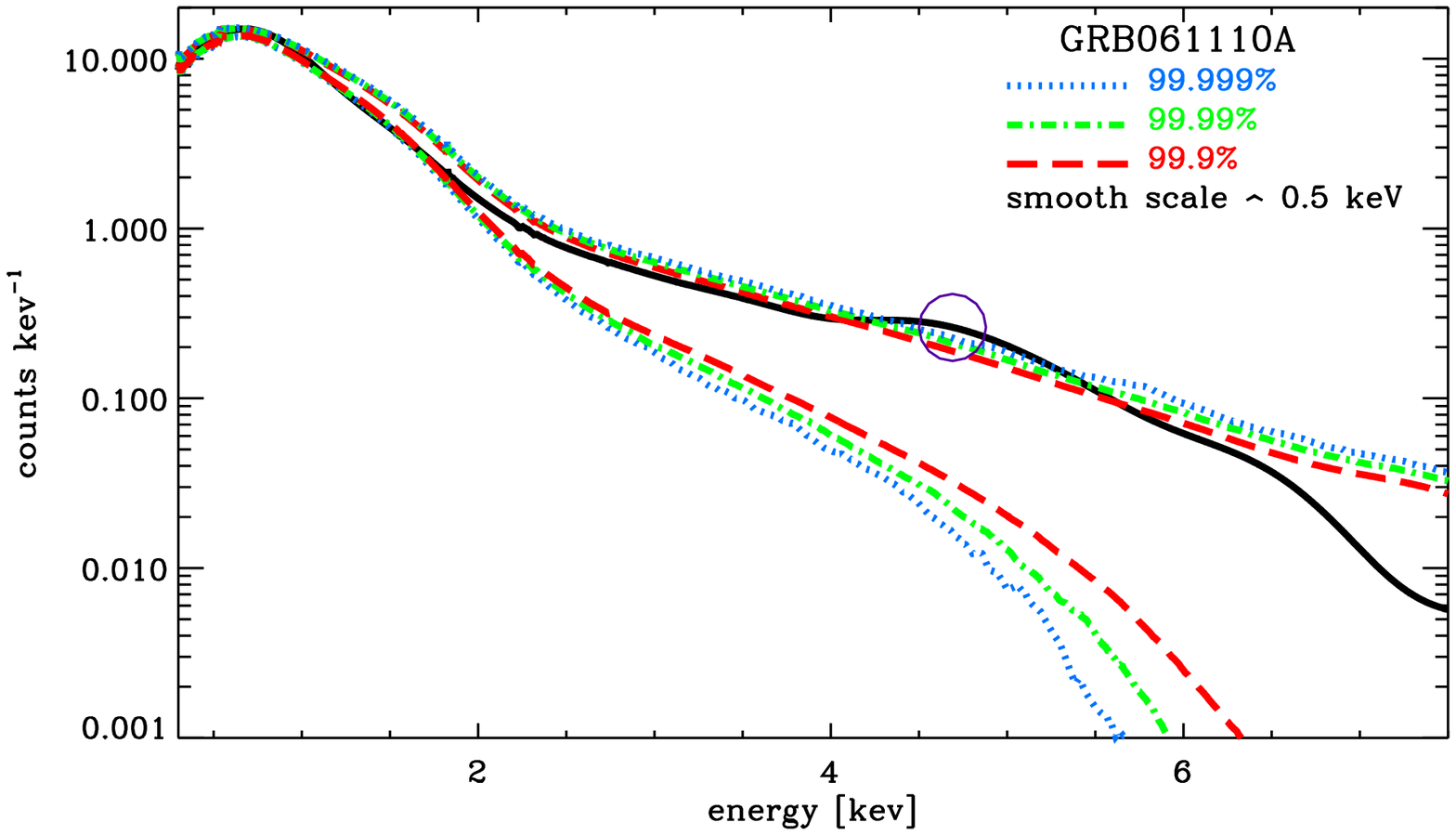} \\  
\end{tabular}
\caption{Data are compared with the single trial confidence contours
as resulted from MC simulations (10$^6$ tests for each GRB; see
Tab.\ref{tab:tab1}).  In the plot, data and simulations contours are
both smoothed to the energy resolution scale which maximizes the
spectral feature signal. These are 7, 10, 3 and 5 times the
instrumental spectral resolution, approximately corresponding to 0.7,
1.0, 0.3, 0.5 keV respectively for the four afterglows (note that
energy resolution varies within the XRT energy band so these are only
rough estimates). The circles indicate where the most significant
excess is detected.}
\label{fig:f5}
\end{figure*}

In order to improve the accuracy in the calculation of the excess
statistical significance, for the sample of four spectra we enlarged
the simulated sample to 1,000,000. Results are reported in
Table~\ref{tab:tab1} and illustrated in Fig.~\ref{fig:f5}.  We note
that in the case of GRB 060729, with 1,000,000 MC tests we obtained a
significance value slightly lower than the one we obtained with
100,000 tests and slightly lower than the 99.99\% threshold
(99.9875\%). Because the two results are perfectly consistent (within
$1\sigma$ errors) we kept GRB 060729 in our sample.
To calculate the corresponding multi-trial significance, we took into
account that we searched for excesses on a sample of 13 spectra on
five different energy resolution scales. For each energy scale we
considered a different number of energy resolution elements:
40,14,8,4,3 respectively for the 1,3,5,10,16 scales. The results are
reported in Table~\ref{tab:tab1}.  Because we searched for features on
a homogeneous sample of 13 spectra, resulting in 4 successes, we can
also estimate the joint statistical significance. In particular, we
set a lower limit to the total joint probability using the binomial
distribution and assuming that all the four spectra have the same
multi-trial significance equal to the lowest (P=99.04\%).  Then we
searched for the probability to have a rate of 4 successes out of 13
tests with mean probability P, resulting in a joint significance of
99.9994\% ($4.3 \sigma$) which ensures that the features we detected
cannot be explained as statistical fluctuations beyond any reasonable
doubt.
\begin{table*}
\begin{center}
\caption[]{ (I)GRB: Name of the GRB; (II)Model: fitting model (see
text); (III)$N_{{\rm H},z}$: neutral absorber column density at the
source redshift; (IV)$\Gamma_1$, ($\Gamma_2$): photon index of the
first (second, present if model is DPL) power law; second photon index
is fixed to the value of the late afterglow; (V)E: rest frame
blackbody temperature if Model is BB1, or BB2 or mean value of the
Gaussian line if model is GAU; (VI)R$_b$: blackbody radius;
(VII)$\sigma_g$: $\sigma$ of the Gaussian line; (VIII)L: unabsorbed
luminosity of the extra component in the rest--frame [0.3-10] keV energy
band; (IX)$\chi^2$(dof): reduced $\chi^2$ (degrees of freedom) in two
different energy bands.}
\begin{tabular}{l|c|clccccc} 
\hline
\hline
GRB          &Mod&$N_{{\rm H},z}$&$\Gamma_1$, ($\Gamma_2$)&E&R$_b$&$\sigma_g$&L&$\chi^2$(dof)\\
             &   &$ 10^{22}$cm$^{-2}$&&keV&cm&keV&10$^{47}$erg s$^{-1}$&0.3-10keV, 1-10keV\\
\hline	     				  			    				 			    	 			    										 									 
060502A&SPL&0.63$_{-0.11}^{+ 0.12}$&3.39$_{-0.16}^{+ 0.18}$        &--                     &--                           &--                  &575.$\pm$33.0&1.29(36), 2.03(11)  \\
       &DPL&1.26$_{-0.21}^{+ 0.23}$&4.97$_{-0.42}^{+ 0.44}$, (1.99)&--                     &--                           &--                  &39.8$\pm$8.50&0.81(35), 0.61(10)  \\
       &BB1&0.41$_{-0.16}^{+ 0.18}$&2.22$_{-0.28}^{+ 0.28}$        &0.33$_{-0.03}^{+ 0.03}$&7.3$_{-0.8}^{+ 2.1}$10$^{12}$&--                  &83.1$\pm$11.1&0.72(34), 0.70(9)   \\
       &BB2&1.02$_{-0.17}^{+ 0.19}$&4.16$_{-0.30}^{+ 0.33}$        &3.24$_{-0.80}^{+ 2.27}$&2.2$_{-0.7}^{+ 1.9}$10$^{10}$&--                  &7.23$\pm$1.45&0.80(34), 0.72(9)   \\   
       &GAU&1.04$_{-0.18}^{+ 0.17}$&4.20$_{-0.33}^{+ 0.27}$        &1.03$_{-1.03}^{+ 8.97}$&--                           &8.8$_{-4.8}^{+ 0.2}$&7.88$\pm$1.61&0.82(33), 0.80(8)   \\
\hline 		 	    		            		       	 			     	 		 	     	      	  	           
060729&SPL&0.22$_{-0.02}^{+ 0.02}$&3.91$_{-0.10}^{+ 0.10}$        &--                     &--                           &--                  &77.2$\pm$2.50&1.24(105), 1.60(38)\\
      &DPL&0.32$_{-0.03}^{+ 0.03}$&4.63$_{-0.19}^{+ 0.20}$, (1.94)&--                     &--                           &--                  &2.28$\pm$0.39&1.05(104), 1.14(37)\\
      &BB1&0.07$_{-0.04}^{+ 0.04}$&2.94$_{-0.20}^{+ 0.20}$        &0.20$_{-0.02}^{+ 0.02}$&6.4$_{-0.8}^{+ 5.9}$10$^{12}$&--                  &8.62$\pm$0.84&1.01(103), 1.08(36)\\
      &BB2&0.32$_{-0.03}^{+ 0.04}$&4.54$_{-0.19}^{+ 0.21}$        &1.25$_{-0.15}^{+ 0.22}$&5.6$_{-0.9}^{+ 1.2}$10$^{10}$&--                  &1.01$\pm$0.16&1.03(103), 1.08(36)\\
      &GAU&0.30$_{-0.03}^{+ 0.04}$&4.40$_{-0.19}^{+ 0.25}$        &3.75$_{-2.75}^{+ 1.07}$&--                           &2.0$_{-0.6}^{+ 1.4}$&0.76$\pm$0.12&1.03(102), 1.09(35)\\
\hline 		 	    		            		       	 			     	 		 	     	      	  	      
060904B&SPL&0.69$_{-0.04}^{+ 0.04}$&3.55$_{-0.08}^{+ 0.08}$        &--                     &--                           &--                  &329.$\pm$0.08&1.00(103), 1.12(60)\\
       &DPL&0.86$_{-0.07}^{+ 0.08}$&4.22$_{-0.23}^{+ 0.23}$, (2.15)&--                     &--                           &--                  &15.2$\pm$2.90&0.92(102), 0.98(59)\\
       &BB1&0.50$_{-0.09}^{+ 0.10}$&2.91$_{-0.26}^{+ 0.24}$        &0.30$_{-0.03}^{+ 0.03}$&4.9$_{-0.6}^{+ 0.6}$10$^{12}$&--                  &25.2$\pm$3.50&0.93(101), 0.97(58)\\
       &BB2&0.79$_{-0.06}^{+ 0.06}$&3.84$_{-0.14}^{+ 0.16}$        &3.02$_{-1.03}^{+ 5.98}$&1.4$_{-0.6}^{+ 3.5}$10$^{10}$&--                  &2.22$\pm$0.39&0.91(101), 0.96(58)\\
       &GAU&0.74$_{-0.05}^{+ 0.05}$&3.67$_{-0.09}^{+ 0.10}$        &7.85$_{-0.25}^{+ 0.16}$&--                           &0.5$_{-0.2}^{+ 0.3}$&1.10$\pm$0.18&0.87(100), 0.90(57)\\
\hline 		 	    		            		       	 			     	 		 	     	      	  	      
061110A&SPL&0.13$_{-0.03}^{+ 0.04}$&3.32$_{-0.12}^{+ 0.13}$        &--                     &--                           &--                  &40.6$\pm$1.60&1.74(77), 2.54(26)  \\
       &DPL&0.45$_{-0.07}^{+ 0.07}$&4.99$_{-0.29}^{+ 0.30}$, (1.75)&--                     &--                           &--                  &4.23$\pm$0.59&1.16(76), 0.99(25)  \\
       &BB1&0.03$_{-0.03}^{+ 0.05}$&2.17$_{-0.17}^{+ 0.17}$        &0.23$_{-0.02}^{+ 0.01}$&5.1$_{-0.3}^{+ 0.6}$10$^{12}$&--                  &9.49$\pm$0.82&1.07(75), 1.07(24)  \\
       &BB2&0.37$_{-0.06}^{+ 0.06}$&4.40$_{-0.25}^{+ 0.27}$        &1.82$_{-0.25}^{+ 0.36}$&3.6$_{-0.6}^{+ 0.9}$10$^{10}$&--                  &1.86$\pm$0.25&1.18(75), 1.24(24)  \\ 
       &GAU&0.34$_{-0.04}^{+ 0.07}$&4.29$_{-0.17}^{+ 0.26}$        &1.00$_{-0.25}^{+ 0.36}$&--                           &5.2$_{-1.0}^{+ 0.8}$&1.89$\pm$0.26&1.20(74), 1.33(23)  \\
\hline
\end{tabular}
\label{tab:tab2}  
\end{center}
\end{table*}
\subsection{Instrumental effects}
We investigated the possibility that these features are produced by
some instrumental effects such as pile--up. This cannot be the case,
because the mean count rate in all the four cases is less than 33
counts s$^{-1}$ and it is always below 70 counts s$^{-1}$, a factor 3
below the pile--up threshold in WT ($\sim$ 200 counts s$^{-1}$, see
Campana et al. 2008).  We can also exclude that these features are
produced by some anomalous hot pixels from the comparison with the
expected point spread function (\cite{Moretti05}).

We can also exclude that the excesses are due to 
uncertainties in the instrument response calibration.  All the
excesses are found at energies $>$ 3.5 keV (Table~\ref{tab:tab1}) in
the central part of the CCD ($<$ 70 pixels, equivalent to $\sim$ 3
arcmin).  In this position and energy range the systematics in the
effective area, quantum efficiency and energy redistribution
calibrations are less than 5\%.
(see\footnote{http://heasarc.gsfc.nasa.gov/docs/heasarc/caldb/swift/docs/xrt}
and Godet et al. 2007b). In fact all the major instrumental edges are
below 3.5 keV (Au for the mirror, Al for the filter, C, N, O and Si
for the CCD) and there is no evidence of position dependence at
energies higher than 0.5 keV.

The excesses cannot result from the background subtraction procedure.
In fact, the expected background events, in the extraction region we
considered, in 100 second exposure, is 1.2$\pm$0.2; among them
0.6 are expected with energy higher than 2 keV (Moretti et al. 2007).
Moreover, we checked our data against uncovered anomalies performing
our analysis with different background extraction regions and we found
perfectly consistent results.

In the case of GRB060904B, a galaxy cluster with a core radius of 12
arcsec is present in foreground, at 2.3 arcminutes of projected
distance from the GRB. By modeling the cluster surface brightness with
a King profile in the extraction region of the afterglow we expect to
find $<$1 photon (99.99\% confidence) from the cluster in the 80 s
exposure.
\section{Modeling the data with an extra component \label{sect:extra}}
\begin{figure*}
\begin{tabular}{cc}
\includegraphics[width=9cm]{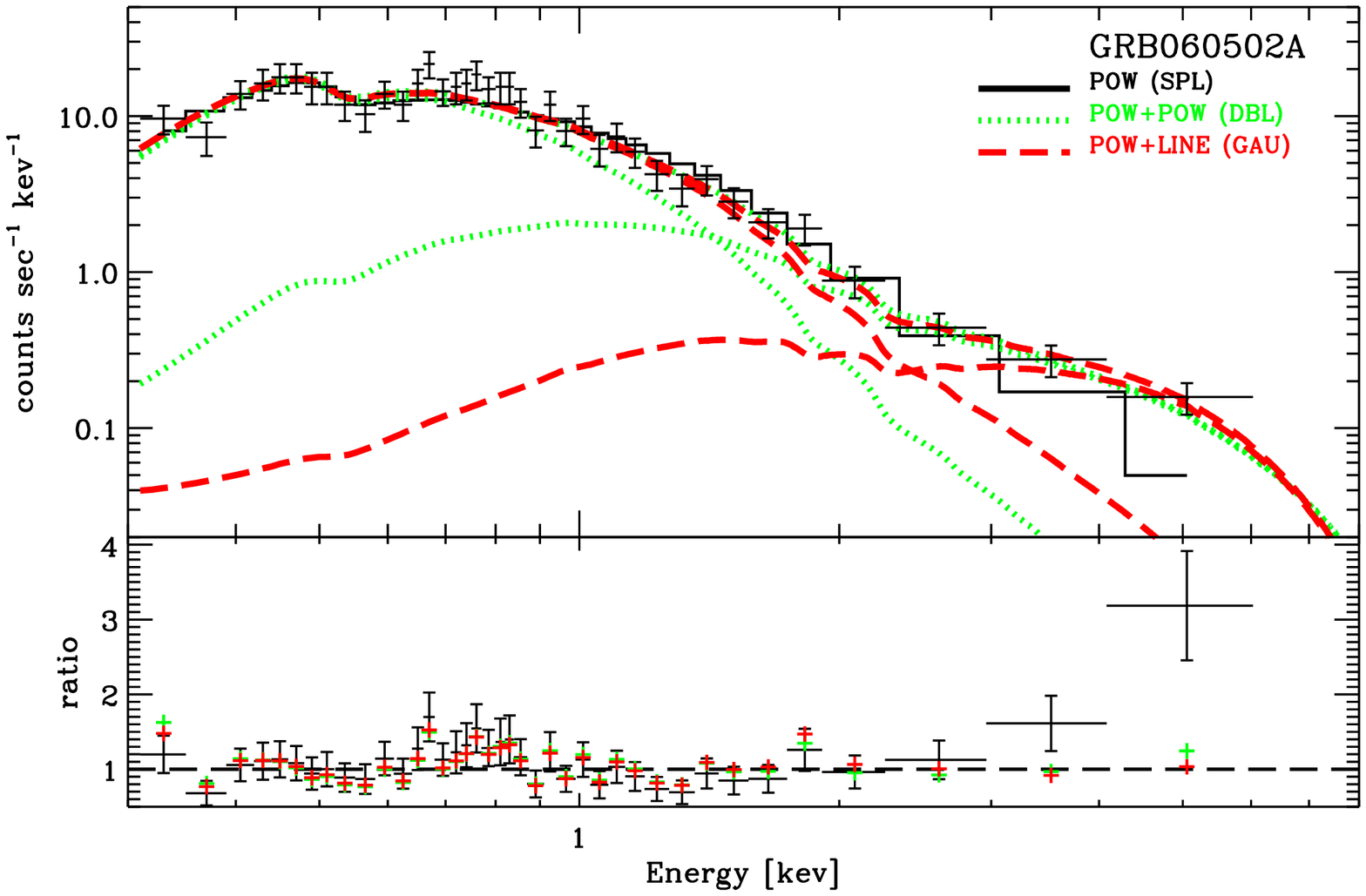} & \includegraphics[width=9cm] {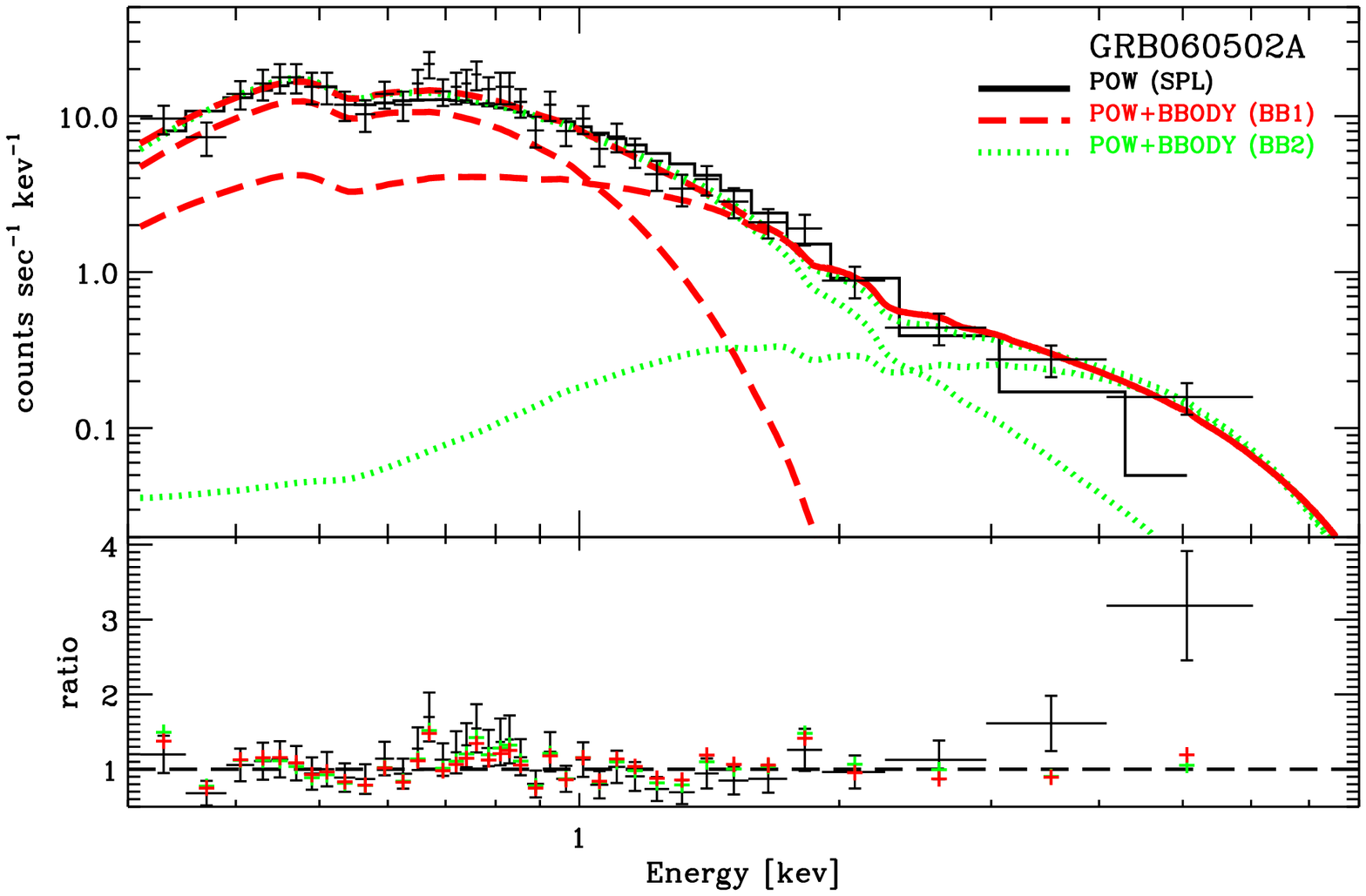} \\ 
\includegraphics[width=9cm]{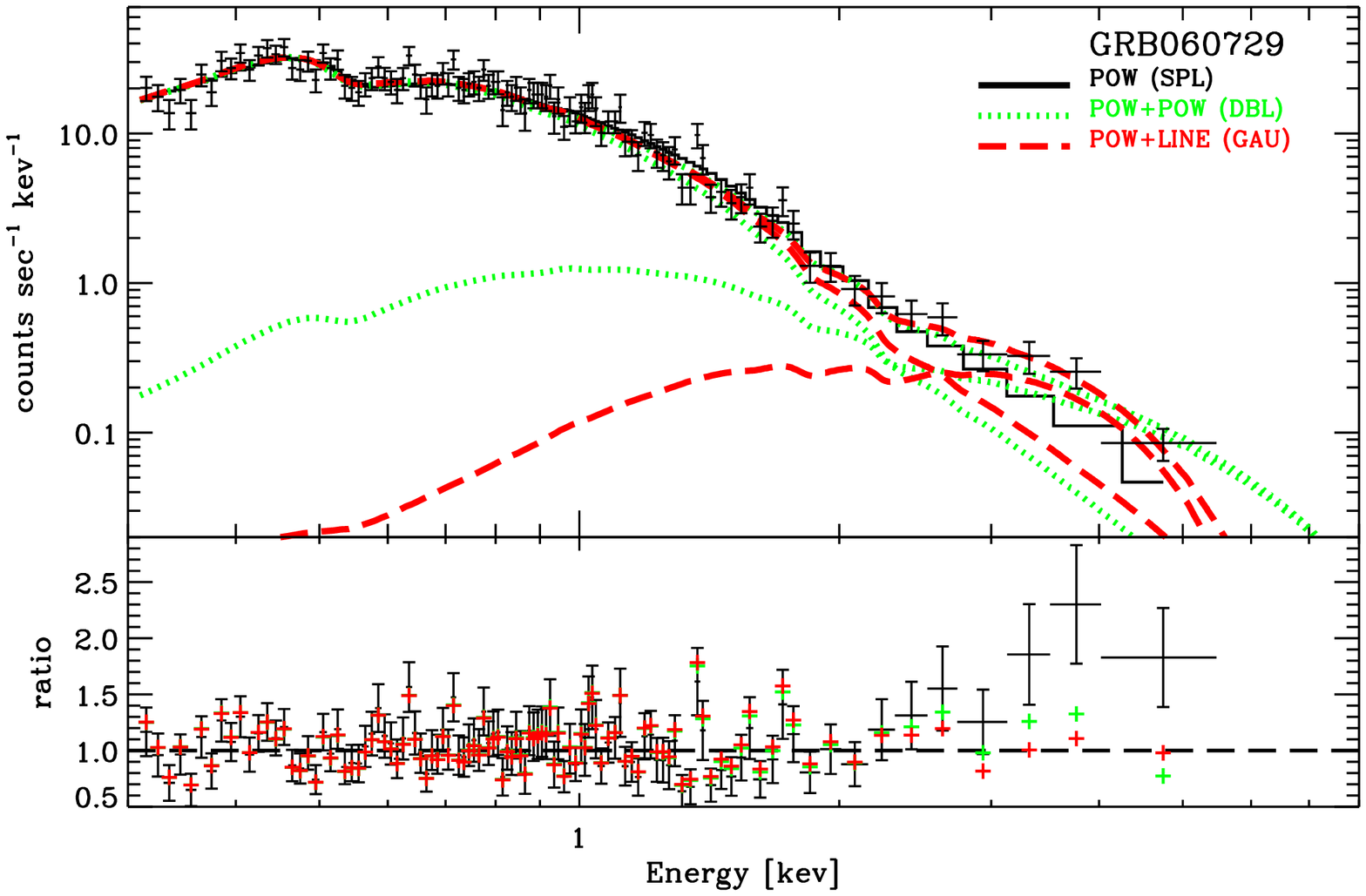}  & \includegraphics[width=9cm]{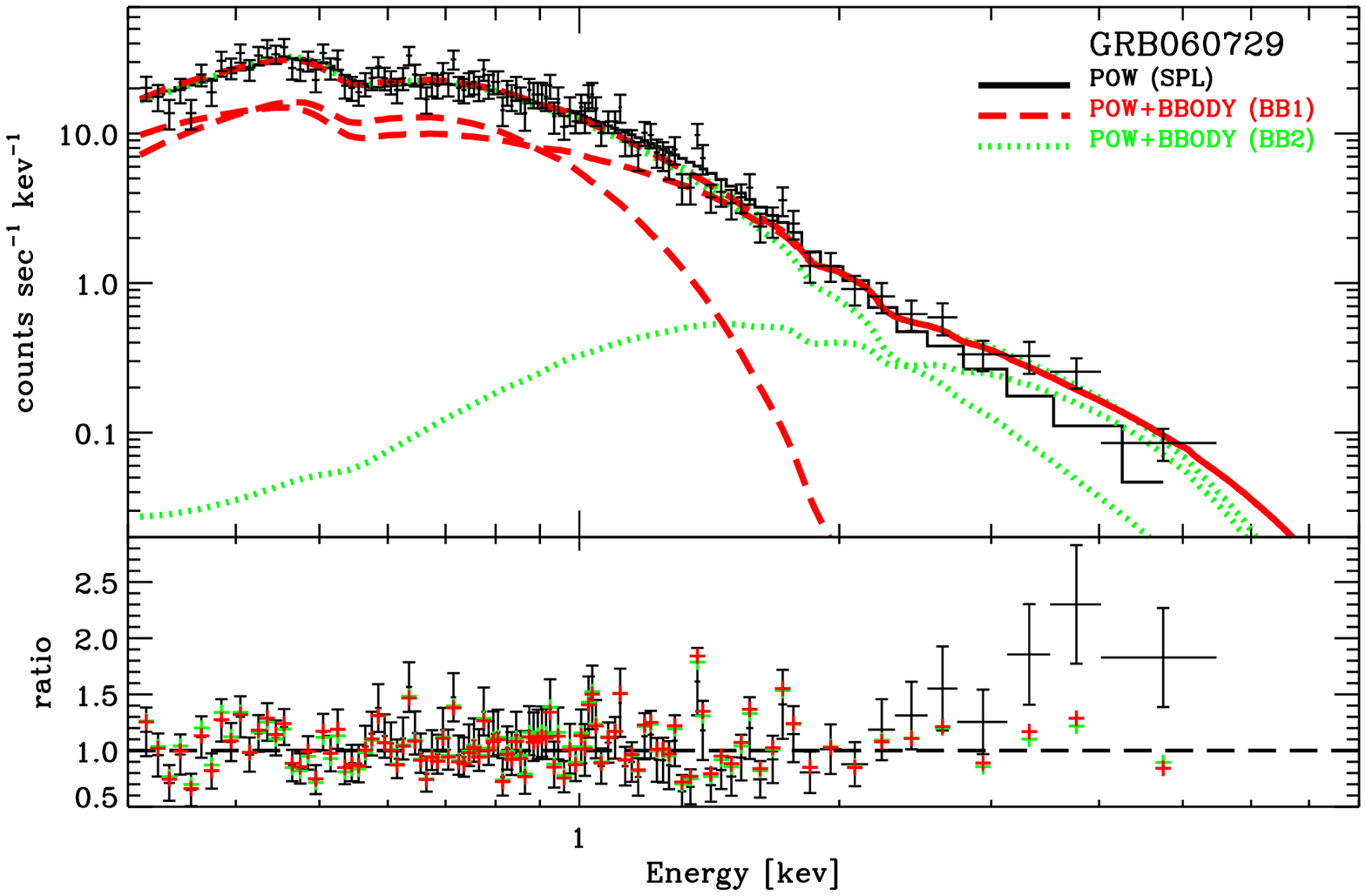} \\  
\end{tabular}
\caption{The anomalous spectra in the GRB 060502A and GRB060729
afterglows. For the sake of clarity, we use two plots for each
spectrum: we plot DPL (dotted grey line) and GAU (dashed line) on the
left plots and BB1 (dashed line) and BB2 (dotted grey line) on the
right plots. In each plot we show the contribution of the single
components and their sum, with the same line style and color. SPL best
models are plotted with the continuous black step-line.  In the lower
panels of each plot we show the ratio between data and models with the
same color code. For the sake of clarity we plot errors only for SPL
models}
\label{fig:f6}
\end{figure*}
As discussed in the previous section we found a sub-sample of four
spectra, belonging to four different GRB which present highly
significant deviations from the SPL models.  In fact, if fitted with SPL,
they all four present a clear excess at high energies.

As already said, most of the departures from SPL models found in time
resolved spectral analysis of Swift observations of early afterglow
have been explained so far by the curvature of the spectrum and the
presence of the $\nu F_{\nu}$ peak (E$_{\rm{peak}}$) within the XRT
band (Falcone et al. 2006, Butler \& Kocevski 2007, Goad et al. 2007,
Mangano et al. 2007).  The four spectra we are considering can be
modeled neither by a Band model (Band et al. 1993) nor by a power law
with an exponential cutoff.  In the time intervals we are considering,
the BAT signal is almost null. This prevented us from studying the
spectrum in the combined energy band. However, any attempt at fitting
the XRT spectra with a cutoff power law or with a Band model could not
constrain the E$_{\rm peak}$ within the XRT band and therefore did not
improve the SPL fit.  We found that our analysis is consistent with
previous papers. In particular Butler \& Kocevski (2007) performed
time resolved spectroscopic analysis of BAT and XRT simultaneous
observations of a large sample of Swift GRBs. For GRB 060729 and GRB
060904B, in particular, they found that, during the early phases of
the afterglow, the energy peak (E$_{\rm peak}$) of the prompt emission
spectrum transits in the X-ray band.  The time intervals where we
observe excesses in the spectrum of GRB 060904B (Table~\ref{tab:tab1})
roughly correspond to the last 5 temporal bins of their analysis
(240-315 seconds in the observer frame).  Although they do not
explicitly report the best fit parameter and the $\chi^2$ values, it
is clear from their Fig. 9 (third panel from the top on the right)
that while the Band model is a good description of the data at the
beginning of the XRT observation, in these five particular intervals,
Band model gives a very poor description of the data, always leaving
one parameter unconstrained.  The same conclusion can be also drawn
for GRB 060729, where our time interval corresponds to their last
three time slices. We also note that for this GRB the same conclusion
is also confirmed by Grupe et al. (2007) who found that cutoff power
law model gives $\chi^2_\nu$ values larger than 1.5 in the same time
interval.

\begin{figure*}
\begin{tabular}{cc}
\includegraphics[width=9cm]{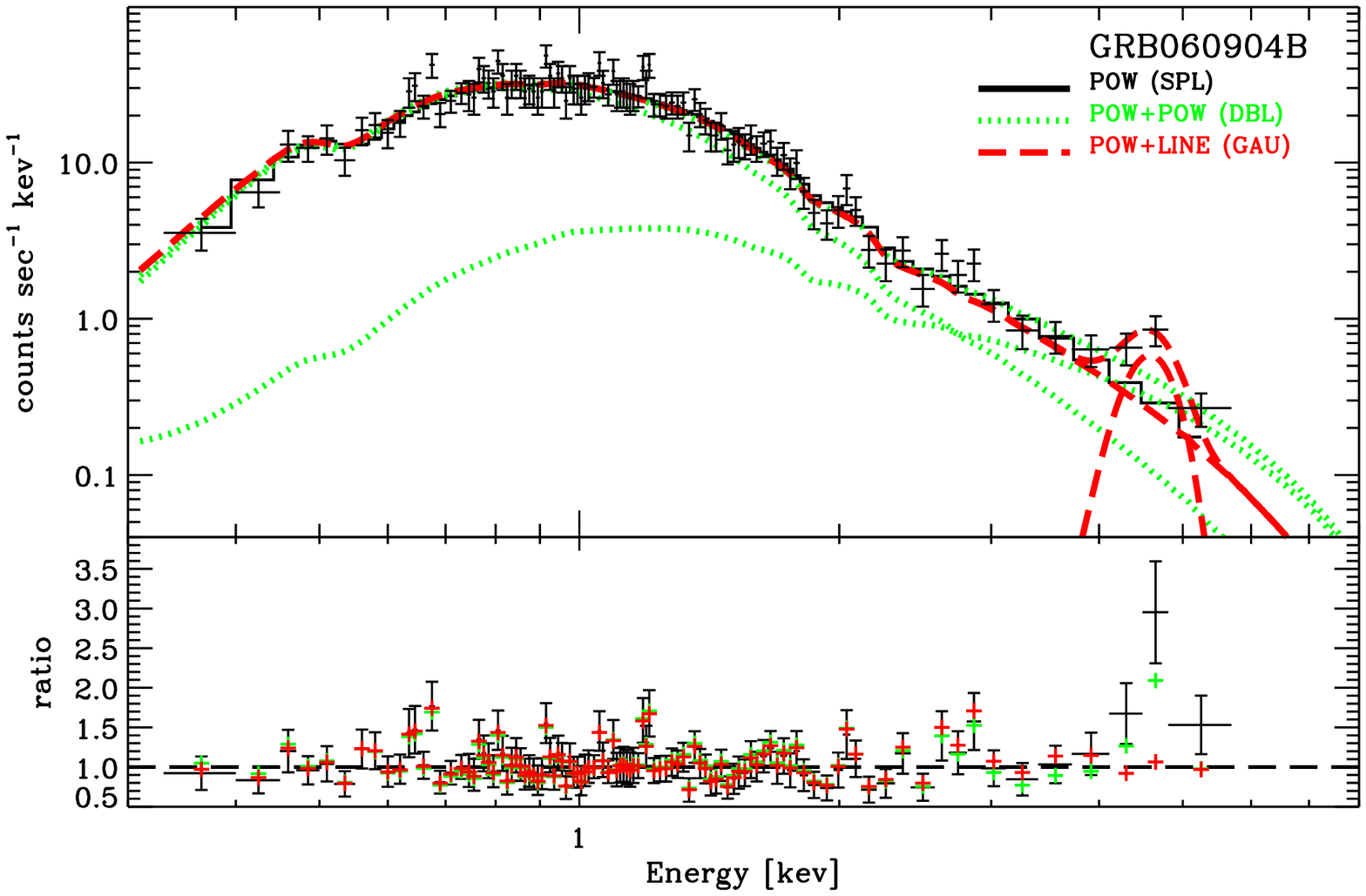} & \includegraphics[width=9cm]{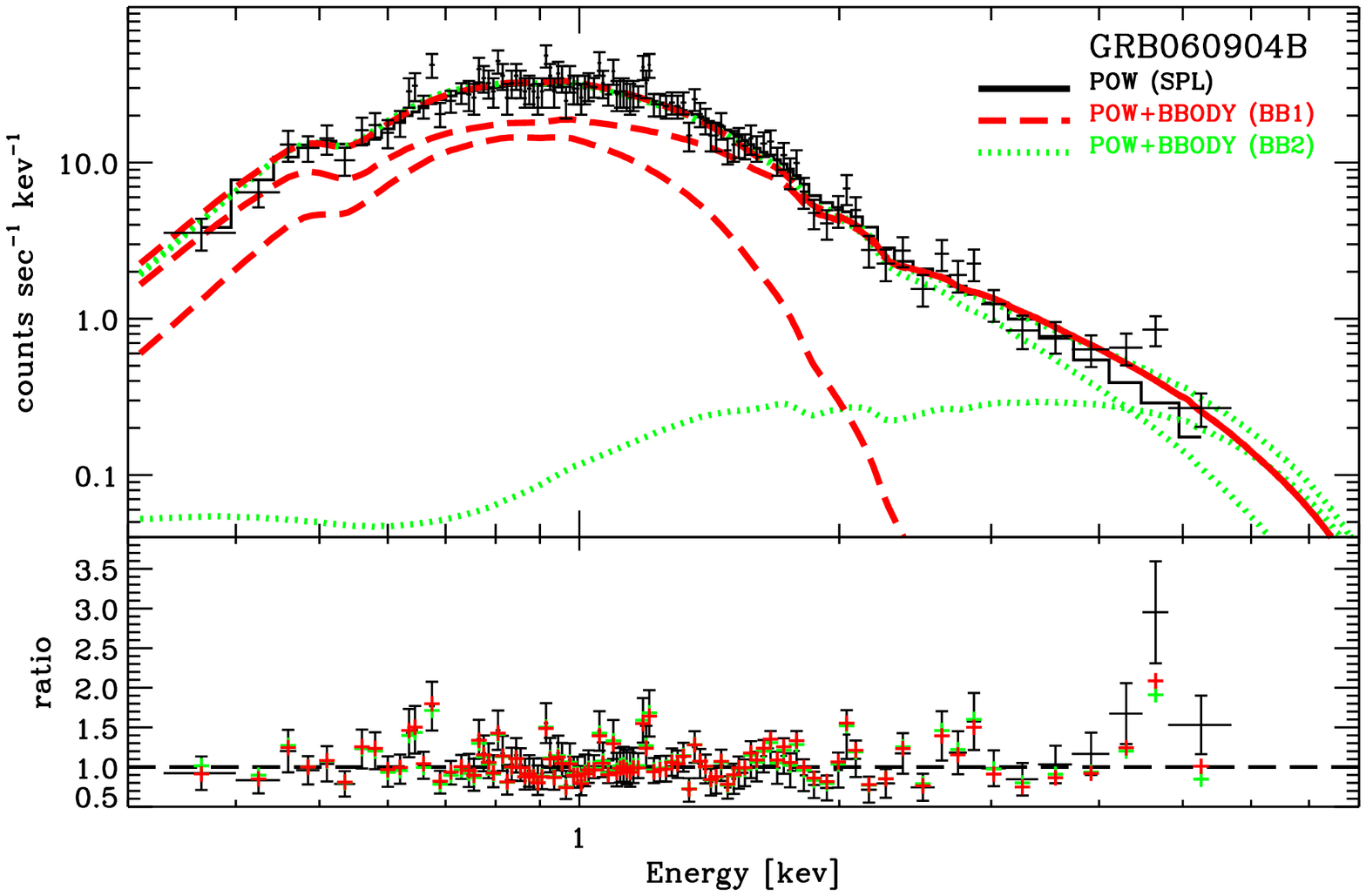} \\ 
\includegraphics[width=9cm]{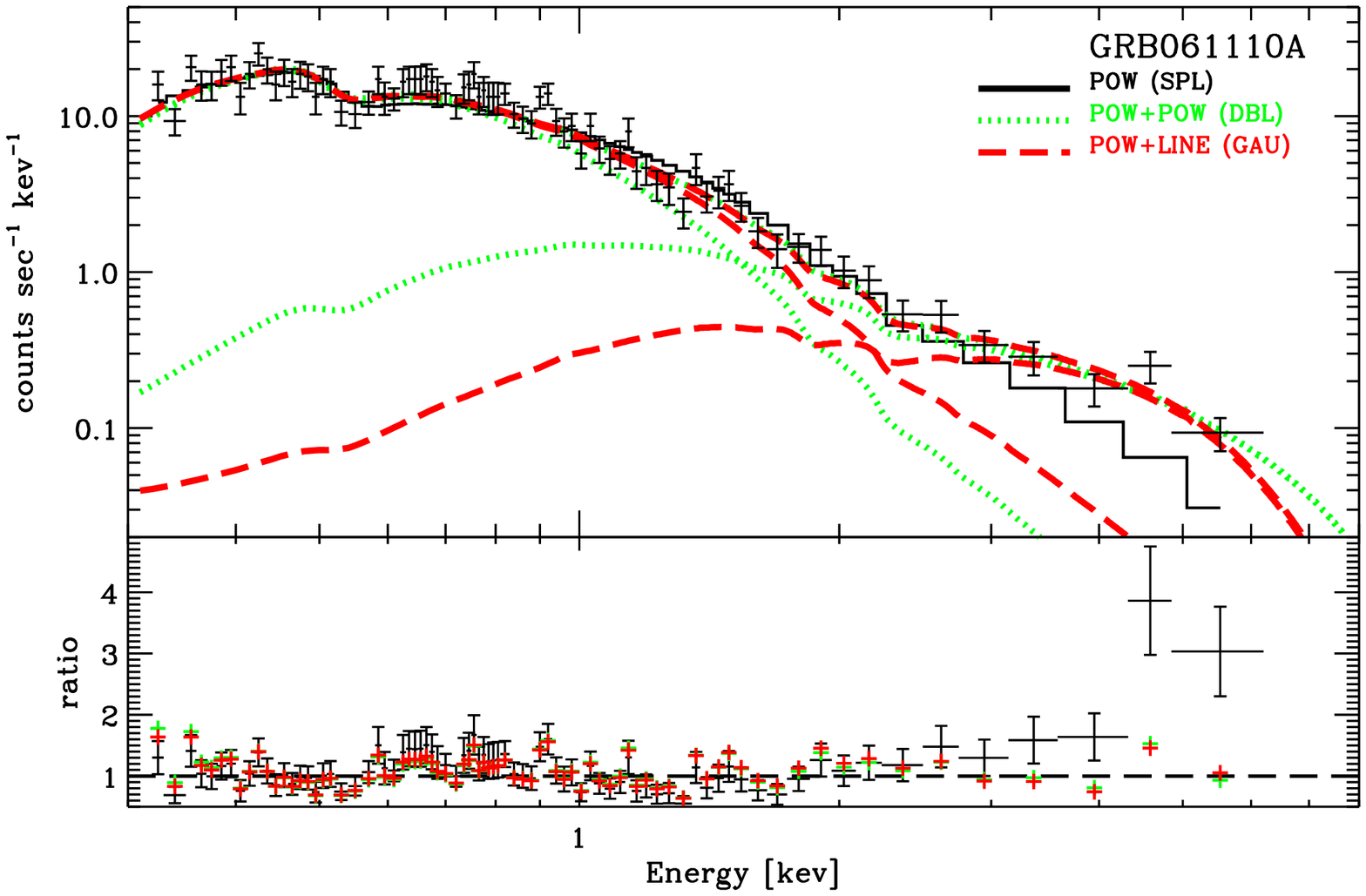} & \includegraphics[width=9cm]{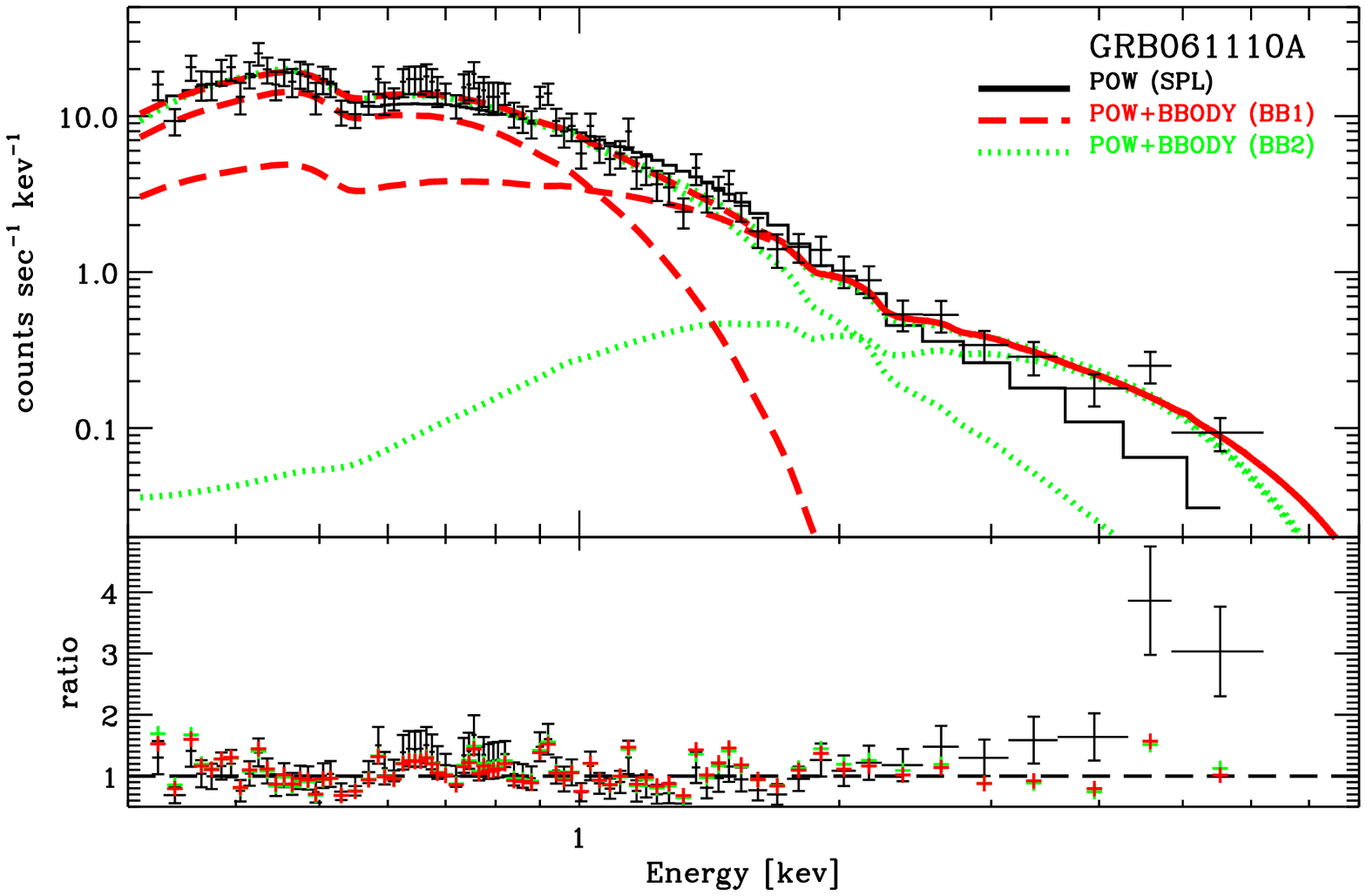} \\  
\end{tabular}
\caption{Same of Fig.~\ref{fig:f6}, for GRB 060904B and GRB061110A afterglow anomalous spectra.}
\label{fig:f7}
\end{figure*}

Therefore we tried to fit the data adding three different components
to the SPL: (i) a second power law component with the slope frozen to
the value of the late afterglow spectrum letting only the
normalization vary (DPL); (ii) a blackbody; (iii) a Gaussian line
(GAU).  As it will be explained later (Sect.~\ref{sect:s4.2}), we
found that, with the blackbody model, two equally good fits could be
found with quite distinct parameter values. Therefore we considered a
blackbody with temperature varying in the 0.1-10 keV energy band with
initial guess kT=0.2 keV (BB1) and kT=2 keV (BB2).
\footnote{Note that with DPL, BB1, BB2 and GAU we intend the model SPL
plus the extra component.}.  The results are reported in
Table~\ref{tab:tab2} and shown in (Fig.~\ref{fig:f6}-\ref{fig:f7}).
\section {Discussion}
\subsection{Energetics and time variability of the spectral features \label{sect:uppe}}
The DPL, BB2 provided an extra component to SPL which compensates the
high energy residuals. In BB1 models, the blackbody component
represents a significant fraction of the softer part of the spectrum,
while the excesses at high energies are accounted for by the power law
component.  The (unabsorbed) luminosities of the additional components
calculated in the [0.3-10] keV rest--frame energy band are typically
1--10\% of the total, in DPL, BB2 and GAU models, while it is 10--20\%
of the total in BB2 models (Table~\ref{tab:tab2} and
(Fig.~\ref{fig:f6}-\ref{fig:f7})).  As illustrated in
Fig.~\ref{fig:f4} for all the four afterglows we could perform time
resolved spectroscopy. The time intervals from which we extracted the
anomalous spectra correspond to the last WT spectrum, right before the
canonical steep--shallow light--curve break (\cite{Nousek06}).  With
the same criteria adopted for the original 13 very soft spectra we did
not find any other significant deviation from SPL model in any of the
time slices considered.  In order to study the time variability of the
spectral features, for each of the four GRBs, we estimated the
detection upper limits.  To do this, for each slice, we added an extra
component (second power law, blackbody and Gaussian) with the
parameter value set to the best DPL, BB1, BB2 and GAU values, letting
the normalization free to vary. We set the upper limits for the
detection when these additional components produce a factor of 3
worsening in terms of null hypothesis probability of the $\chi^2_\nu$
of the SPL fit.  We note that this should be considered as a rough
estimate of the upper limits; a rigorous calculation of all the upper
limits would have required an unrealizable number of
simulations. However we verified that, at least in one case, the upper
limits roughly calculated differs by less than 30\% from the one
rigorously calculated.

The upper limits vary during the afterglow depending on the flux and
on the softness of the spectrum. In particular, as shown in
Fig.~\ref{fig:f4}, the detection of the spectral features, in the last
spectrum before the plateau phase, coincides with a drop of the
detection threshold.  This is due to the simultaneous flux decay and
spectral softening which allow the detection of spectral features in
the higher part of the energy band.  We note that GRB 060904B did not
show this component at early times, although the sensitivity was also
good then.  We also note that in at least three cases the extra
components present in the last part of the light curve steep decay
disappear in the shallow phase, although the upper limits in this time
slices are very low (in the case of GRB 060904B, the observation
stopped during the steep decay of a giant X--ray flares). Evidently,
whatever its nature, the emission mechanism responsible for these
spectral features varies on a time scale similar to the prompt
emission.
\subsection{Double power law (DPL)}
In three cases DPL model provides the best improvement in the fit,
taking into account that we add only one extra parameter to the SPL
model. This model would provide a natural explanation to the excesses
we observe. In fact, because the afterglow spectrum is significantly
harder than the prompt tails, when the prompt flux decreases and
softens, the afterglow emission becomes visible at higher
energies. This would easily explain the fact that the excesses are
detected just before the steep-shallow light--curve break and
disappear in the following time slice.  On the other hand, as shown in
Fig.~\ref{fig:f4}, the excess luminosity is much larger than the
expected contribution of the afterglow forward shock component alone
(e.g. \cite{Sari97}, Willingale et al. 2007).  But it might be
explained by the radiation produced by an extreme reverse shock in the
X--ray band (see Zhang et al. 2006, Kobayashi \& Zhang 2007).
\subsection{Power law plus blackbody (BB1, BB2)  \label{sect:s4.2}}
\begin{figure}
\includegraphics[width=9cm]{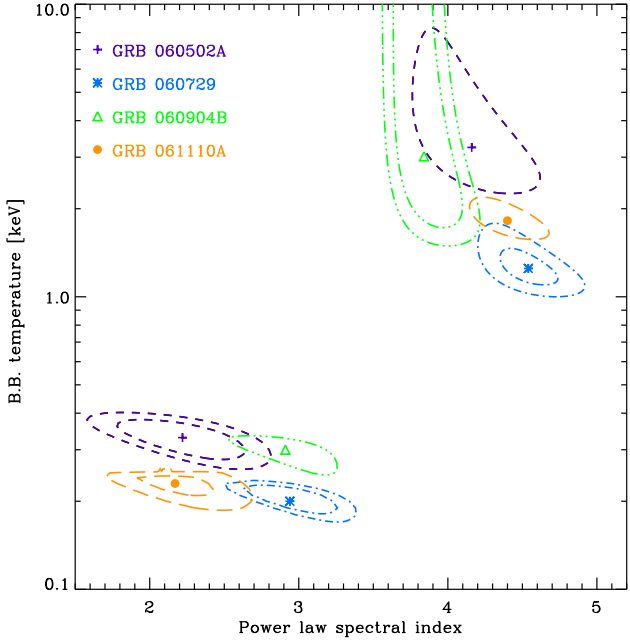} 
\caption{Confidence contours (68\% and 90\%) of the single power law+blackbody model for the 
four different GRBs.}
\label{fig:f8}
\end{figure}

Good fits are also provided by adding to the SPL a blackbody component
(2 extra parameters). In all the four afterglows, with this model, the
fit results depend on the blackbody temperature initial guess. As
shown in Fig.~\ref{fig:f8}, $\chi^2$ has two (and only two) different
but equally significant minima. BB1, with initial guess kT=0.2 keV,
gives good fits with (redshift-corrected) temperatures in the range
0.20--0.33 keV, radii in the range (4.9--7.3)$\times$10$^{12}$ cm and
power law indexes in the range 2.2--2.9. BB2, with initial guess kT=2
keV gives good fits with red--shift corrected temperatures in the
range 1.3--3.2 keV, radii in the range (1.4--5.6)$\times$10$^{10}$ cm
and power law indexes in the range 3.8--4.5. For GRB 060502A and GRB
061110A, BB1 fits provided slightly better $\chi^2$ as calculated on
the whole energy band, while for GRB 060729 and GRB 060904B the fits
yield equivalent results. Interestingly, in each of the four cases,
the $\chi^2 $ matrix projection on the photon index -- temperature
parameter plane has always two well defined minima which are split in
two different regions of the plane (Fig.~\ref{fig:f8}).  This explains
why fit results depend on the initial guess value of the blackbody
temperature.

In the case of GRB 060729 our BB1 model is consistent with the results
of Grupe et al. (2007) and Godet (2007a). They interpreted the thermal
emission as due to the photospheric emission from X--ray flares.  In
fact, assuming that X--ray flares, in the early afterglow, are
produced by the same mechanism of the prompt phase, a thermal
component in the early afterglow spectra is expected in a similar way
to the prompt emission (Ryde et al 2006). This interpretation can be
easily extended to the other three GRBs that have very similar
characteristics.
The prompt thermal emission described by BB2 model can also be
explained as a shock break--out. The shock--heated plasma would be at
temperatures of (1.4--3.7) $\times$ 10$^7$ K, with a luminosity of
(1.0--7.0)$\times$10$^{47}$ erg s$^{-1}$ corresponding to a radius of
(1.4--5.6)$\times$10$^{10}$ cm.  Assuming the duration of the time
slice in which we extracted the spectrum as the duration of the
emission, we obtain a total energy of (1.0--1.8)$\times$10$^{49}$ erg.
The energy, variability, luminosity and temperature we observe in the
detected excesses are consistent with the characteristics of the
transient event from shock breakout in Type Ibc supernovae, produced
by the core-collapse of WR stars surrounded by dense winds
(\cite{Li07}).
\subsection{Power law plus a Gaussian line (GAU)}
Adding a Gaussian line to the SPL in three cases does not improve the
fit with respect to BB2, although it uses one extra parameter. In
fact, for GRB060502A, 060729 and 061110A the best fit is given by low
energy and very broad lines (1-3 keV with $\sigma$ = 2-8 keV), with
best fit values poorly constrained.  The case of GRB060904B is
different and much more intriguing: here the Gaussian fit provides
very well constrained values for the line. In the rest frame the mean
value is 7.85$_{-0.25}^{+ 0.16}$ keV, the width 0.50$_{-0.17}^{+
0.35}$ keV and its luminosity is (1.10$\pm$0.18$)\times$ 10$^{47}$erg
s$^{-1}$.  Interestingly, the Gaussian component can be explained as a
line emission of highly ionized Nickel (7.81 keV).  We refer the
reader to an accompanying paper (Margutti et al. 2007) for a detailed
discussion of the theoretical implications of the possible detection
of Nickel emission at $\sim$ 200 sec after the onset of the GRB.
\subsection{A Nickel line in the GRB 060904B afterglow spectrum ?}
As we saw in the previous sections, the DPL, BB1 and BB2 models
provided significant improvement in the fits with respect to SPL model
in all four cases. In the GRB 060904B afterglow spectrum these models
left some residuals in the high energy part of the spectrum which can
be fit well only by GAU model (Fig.~\ref{fig:f7}).  Since we cannot
lean only on $\chi^2$ statistics to evaluate the goodness of the DPL,
BB1, BB2 fits, we tested the probability that these residuals are
statistical fluctuations of the DPL, BB1, BB2 models.

To this aim, we adopted the same procedure we previously used to test
the hypothesis that the residuals were fluctuations of a SPL model
(Sect.~\ref{sect:s3}). We only replaced the SPL model with the DPL,
BB1, BB2 as input model for the MC simulations.  When we tested the
GRB 060904B afterglow residual significance as a fluctuation of a SPL
model, we found 4.2 $\sigma$ (i.e. 99.9993\%) as a single trial,
corresponding to 3.2 $\sigma$ (i.e. 99.9461\%) as multi--trial (see
Table~\ref{tab:tab1}). If we test, instead, the possibility that this
is a statistical fluctuation of a more complex spectral model (DPL,
BB1, BB2) the single (multi-) trial statistical significance of this
detection is 2.7 (2.2) $\sigma$, with very small differences among the
three models. This means that the deviation from SPL model that we
observe in GRB 060904B can be described as a Gaussian deviation from a
SPL model at 3.2$\sigma$ or from a two component model at 2.2$\sigma$.
\section{Conclusion}
Our most solid result is that we found a small homogeneous and fairly
defined sample of afterglow spectra (the soft sample) for which
deviations from the SPL spectral model are highly probable.  We
started from an homogeneous sample of bright GRB afterglows with known
redshift and we studied their spectral evolution. We split the data in
different time slices and we focused on the softest spectra. In this
sub-sample at least 4 cases out of total of 13 present highly
significant deviations from the SPL spectral model during the
prompt--afterglow transition phase.  We could firmly exclude that
these excesses can be explained as statistical fluctuations of a SPL
spectrum or some instrumental effects.  We also excluded that data can
be fitted by a Band or a cutoff power law model. We fitted these
spectra adding one of three different trial components to the SPL
model. We did not try to discriminate among these different models on
a purely statistical basis, and we discussed them using the component
time variability and energetics.

In a very recent paper Yonetoku et al. (2007) show that the very
soft spectrum of the early afterglow of GRB 060904A presents a feature
which is very similar to what we described in the present work. They
selected and stacked the data in the time intervals of the GRB 060904A
early afterglow, where the spectral photon index is larger than 4.0
(we note that this GRB is not included in our sample because its
redshift is not known). In a very similar way to our results, they
found that, in this spectrum, the data show a clear hardening break
around 2 keV. This feature leaves a significant excess with respect to
the double power law model above 2 keV.  They conclude that this
spectrum consists of two emission components, the second being
consistent with the spectrum of the late afterglow (our DPL model).
These two (independent) studies represent a direct piece of evidence that
the emission observed in the early phases of the afterglow is composed by
more than one component. Differently from Yonetoku et al. (2007), in
our sample, we showed that, if the second component were the emerging
afterglow emission, at early time this should be much more luminous
than the expectation from the classical afterglow model. We showed
that spectral studies of the prompt--afterglow transition phase can be
the starting point to seperate different emitting components and could
provide useful information in order to better understand the afterglow
light curve complexity.

\begin{acknowledgements}
This work is supported at OAB--INAF by ASI grant I/011/07/0 and and by the Ministry of University and Research of Italy
(PRIN 2005025417).
\end{acknowledgements}
%

\end{document}